\documentclass[a4paper,11pt]{article}
\pdfoutput=1

\usepackage{jheppub,braket,cellspace}
\usepackage{amsmath,mathrsfs,amsbsy}

\usepackage[T1]{fontenc}  
\newcommand{\normord}[1]{:\mathrel{#1}:}

\title{\boldmath Localized thermal states and negative energy}

\author{Felipe Rosso}
\affiliation{Department of Physics and Astronomy,
University of Southern California,
\\Los Angeles, California 90089-0484, USA.}

\emailAdd{felipero@usc.edu}

\abstract{We construct localized states defined in a ball or the half-space of a conformal field theory (CFT) in Minkowski that are thermal with respect to the local modular flow. We compute their energy density at arbitrary temperature for a variety of CFTs, and find values for which it is negative and divergent at the boundary. Despite this singular behavior we show that the energy measured by an observer is consistent with the bounds present in the literature. For holographic CFTs these states are captured by hyperbolic black holes in anti-de Sitter, where the negative energy in field theory amounts to the well known negative mass of the black hole. As a byproduct, we show that the Casini-Huerta-Myers proof of the Ryu-Takayangi holographic entanglement formula for the vacuum reduced to a ball can be naturally extended to include half-space regions.}

\begin{document} 
\maketitle

\section{Introduction}
\label{sec:intro}

The quantization of a classical field theory is a highly non trivial procedure which gives rise to several subtle and interesting phenomena. Entanglement correlations between causally disconnected space-time regions and quantum states with negative energy densities are two illustrative examples of intrinsically quantum features that are not present in the classical regime. To better understand the quantum theory and the quantization procedure itself, the investigation of these aspects is an appropriate starting point that can shed valuable insight.

The study of negative energy in quantum field theories (QFTs) has certainly proven to be an extremely fruitful research area. The derivation of bounds restricting the amount of negative energy measured by an observer \cite{Flanagan:1997gn,Fewster:1998pu,Fewster:1998xn,Pfenning:2001wx,Fewster:2003zn,Yu:2003rd,Fewster:2004nj,Blanco:2017akw} (refered as quantum energy inequalities, see Ref. \cite{Fewster:2012yh} for a review) are of vital importance to the second law of thermodynamics \cite{Ford:1978qya,Ford:1990id} and causality when coupling quantum matter fields to classical gravity \cite{Morris,Ori:1993eh,Alcubierre:1994tu,Natario:2001tk}. Moreover, a different type of energy bound such as the averaged null energy condition (ANEC) and the quantum null energy condition (QNEC) \cite{Bousso:2015mna} have been recently proven for a large class of theories \cite{Faulkner:2016mzt,Hartman:2016lgu,Balakrishnan:2017bjg,Kravchuk:2018htv,Ceyhan:2018zfg} and used to derive a variety of very interesting results \cite{Hofman:2008ar,Buchel:2009sk,deBoer:2009pn,Koeller:2017njr,Leichenauer:2018obf}.

Although there has been a lot of attention towards the derivation and application of energy inequalities, less has been done regarding the explicit construction of negative energy states. This is an important aspect given that it can shed light into the detailed structure of the states themselves. The aim of this paper is to systematically construct negative energy states for a wide range of theories defined in Minkowski space-time in any dimension. By restricting to conformal field theories (CFTs) we are able to obtain a one parameter family of states that interpolate between positive and negative energy density inside a ball and the half-space.

Let us first review the setups in which negative energy states appear, which can be roughly classified in three categories. The first is usually referred as quantum coherence and requires a free theory where the fields admit a mode expansion. By constructing specific superpositions of these modes, such as the vacuum and a two particle state \cite{Kuo:1993if}, or squeezed states \cite{Stoler:1969tq,Kuo:1993if} negative energy densities appear. Although explorations in this direction has been considered in Refs. \cite{Pfenning:1998ua,Vollick:1998sk,Ford:2002kz,Yu:2003nc,Ford:2008zz}, they are by construction restricted to free theories. The second setup arises when quantizing a field theory on a curved background geometry, where the space-time curvature induces some negative Casimir energy density on the states \cite{Elizalde:1993ud,Pfenning:1996tb,Herzog:2013ed,Herzog:2015ioa}. Given that the quantization in curved backgrounds is not always simple, this approach is technically more challenging.

The third approach is the one we take in this work and we refer as \textit{localized states}. Given a region $A$ on a Cauchy surface $\Sigma$ let us assume that the Hilbert space decomposes into a tensor product structure $\mathcal{H}=\mathcal{H}_A\otimes \mathcal{H}_{A^c}$, with $A^c$ the complementary region. We name localized states to any density operator defined in the Hilbert space $\mathcal{H}_A$. In fact, from causality and unitarity these states are defined in the causal domain of the region, $\mathcal{D}_A$.\footnote{The causal domain of the region $A$ is defined as the set of points $p$ in Minkowski for which all timelike or null curves through $p$ necessarily intersect with $A$.} Given that negative energy densities are by definition a local feature of the quantum theory, this is perhaps the most natural approach for constructing negative energy states. Previous work which falls into this category (though sometimes from a very different perspective) are the Casimir energy induced by two parallel plates \cite{Brown:1969na,Casimir:1948dh}, the vacuum of the Rindler region \cite{Candelas:1977zza,Candelas:1978gg} and fields in the presence of moving mirrors \cite{Davies:1976hi,Ford:1982ct,Ford:1990ae,Pfenning:1998ua}.

The localized states we consider in this work can be thought as thermal with respect to the modular flow of the region $\mathcal{D}_A$ (thermal in the sense of the KMS condition \cite{Haag:1992hx}). Though the modular flow can be rigorously defined for any region in the context of the Tomita-Takesaki theory \cite{Haag:1992hx,Witten:2018lha}, we will restrict ourselves to simple setups (with $A$ a ball or the half-space) where the modular flow is local and can be described by a timelike coordinate $s$. In this case, given any local operator $\mathcal{O}(x_A)$ with $x_A\in \mathcal{D}_A$, the modular flow is implemented by the unitary operator $U(s)$ according to
\begin{equation}\label{eq:146}
U(s)\mathcal{O}(x_A)U^\dagger (s)=
  \mathcal{O}(x_A(s))\ ,
\end{equation}
where $x_A(s)\in \mathcal{D}_A$. From this we can define a thermal state with inverse temperature $\beta$ with respect to translations in $s$ and compute its energy density for arbitrary temperatures. As $\beta$ varies we will find that the energy density in the region $\mathcal{D}_A$ interpolates from positive to negative, providing with a one parameter family of negative energy states.

A seemingly disturbing feature of these localized thermal states is the fact that their energy density diverges at the boundary of $\mathcal{D}_A$. In particular, it goes to minus infinity for a given range of $\beta$. The origin of this singular behavior is the fact that the tensor product structure of the Hilbert space $\mathcal{H}=\mathcal{H}_A\otimes \mathcal{H}_{A^c}$ is not rigorously valid in any QFT \cite{Witten:2018lha,Hollands:2017dov}.\footnote{We understand this from a practical level. The entanglement entropy of the Minkowski vacuum reduced to a ball or the half-space can be obtained by integrating the energy density of a localized thermal state \cite{Dowker:2010bu,Casini:2011kv}, where the divergence in the energy density translates in the divergence of the entanglement entropy (which is known to arise from the assumed tensor product structure of the Hilbert space).} To make sure that the states we construct are physical we consider the negative energy measured by an observer restricted to $\mathcal{D}_A$ and compare with several quantum energy inequalities (QEIs) present in the literature. We find perfect agreement with the QEIs, meaning that these localized states are perfectly reasonable states.

One of the most compelling features of these states is the fact that their properties can be mapped through the AdS/CFT correspondence to black holes with hyperbolic horizon.\footnote{The temperature of the localized state is mapped to the horizon temperature of the black hole.} This provides a procedure for studying negative energy states in strongly coupled theories, which is extremely challenging using standard QFT techniques. The negative energy density of the localized state is mapped to a negative mass of the hyperbolic black hole. Though this negative mass has been noted long ago in the literature, its meaning has been poorly understood. From this perspective the negative mass of the black hole is not only natural but expected, given that it arises from the negative energy density of the states in the boundary CFT. This connection has been recently made in Ref. \cite{Rosso:2018yax}, but only for the ball region and the zero temperature case. There it was also suggested that the relation could be generalized to arbitrary temperatures. In this work, we confirm the claims of Ref. \cite{Rosso:2018yax} and make the connection between the localized states and the black holes much more clearer and precise.

In the process of investigating the relation between the localized states and the hyperbolic black holes we stumble upon a very natural extension of the famous construction developed in Ref. \cite{Casini:2011kv}. In that work, the entanglement entropy of the Minkowski vacuum reduced to a ball was mapped to the horizon area of the massless hyperbolic black hole and found to be in perfect agreement with the holographic Ryu-Takayanagi (RT) prescription \cite{Ryu:2006bv,Ryu:2006ef}. In fact, the horizon of the black hole and the RT surface can be shown to be exactly the same, providing with an explicit proof of the RT formula for spherical regions. We find that this proof can be extended beyond spherical regions to include the half-space using the same massless black hole, which allows us to explicitly verify that 
$$S_{\rm EE}=S_{\rm Horizon}=S_{\rm RT}\ ,$$
for the Minkowski vacuum reduced to the half-space of a CFT. Though this relation is certainly not unexpected, it is satisfying to get another setup where the proof can be made explicit. The reader interested in this result and somewhat familiar with the construction of Ref. \cite{Casini:2011kv} is encouraged to go directly to Sec. \ref{sec:massless} and App. \ref{app:entropy}.

The outline of the paper is as follows. We start in the following section by constructing the localized thermal states in the half-space and ball in Minkowski. We explicitly compute their energy density for arbitrary two-dimensional CFTs, massless scalar field and strongly coupled holographic CFTs. Some of the technical calculations, including the quantization of a thermal scalar conformally coupled to a hyperbolic space-time, are relegated to Apps. \ref{app:hyp}, \ref{app:hypeq} and \ref{app:quasi}. In Sec. \ref{sec:Obs} we consider the energy measured by an observer and show that it is consistent with the QEIs present in the literature. For two-dimensional CFTs we show that the localized states saturate the most general inequality in the zero temperature limit. We continue in Sec. \ref{sec:HoloDis} by carefully illustrating the connection between the localized thermal states and the hyperbolic black holes. We finish in Sec. \ref{sec:Dis} with a summary of our results and interesting future directions. The proof of the RT formula for the Minkowski vacuum reduced to the half-space is given in App. \ref{app:entropy}, which also includes some holographic and field theory calculations for the half-space Renyi entropy. In App. \ref{app:Juan} we consider building separable state that are globally defined in the Hilbert space and show that they require an infinite amount of positive energy localized at the entangling surface.

\section{Localized thermal states and energy density}
\label{sec:Loc}

In this section we explicitly construct the localized thermal states in the Hilbert space $\mathcal{H}_A$ with $A$ given by the half-space or a ball in Minkwoski at $t=0$. We compute their energy density at arbitrary temperature for a variety of CFTs. To do so, we must first write the operator $K_s$ which generates the modular flow in Eq. (\ref{eq:146}) and maps $\mathcal{D}_A$ into itself. Given that the flow will be local for the regions we consider, the operator $K_s$ has a simple expression in terms of the stress tensor $T_{\mu \nu}(t,\vec{x})$
\begin{equation}\label{eq:86}
K_{s}=\int_A d\vec{x}\, 
 \zeta^\mu\, T_{\mu t}(t=0,\vec{x})
  \ ,
  \qquad \qquad
  \zeta=
  \frac{\partial}{\partial s}=
  \left.\frac{dx^\mu }{ds}
  \right|_{t=0}\,
  \frac{\partial}{\partial x^\mu}
  \ ,
\end{equation}
where $x^\mu=(t,\vec{x})$. Since this operator plays the role of the Hamiltonian associated to the region $\mathcal{D}_A$, we have a natural definition of a thermal state in $\mathcal{H}_A$ according to
\begin{equation}\label{eq:90}
\rho_A(\beta)=
  \frac{1}{Z}\exp\left(-\beta K_s\right)\ .
\end{equation}
Even though this might not be an ordinary thermal state given that $K_s$ will not always be a conserved quantity, it is thermal in the sense that it satisifes the KMS condition \cite{Haag:1992hx}.

Another crucial feature of taking the regions as the half-space or a ball is that in both cases there is an inverse temperature $\beta=2\pi \ell$ with $\ell$ a length scale, where the thermal state $\rho_A(2\pi \ell)$ is equivalent to the Minkowski vacuum $\ket{0_M}$ of the full Hilbert space $\mathcal{H}$ reduced to $\mathcal{H}_A$, \textit{i.e.}
\begin{equation}\label{eq:85}
\rho_A(2\pi \ell)={\rm Tr}_{A^c}\left(\ket{0_M}\bra{0_M}\right)
  \ ,
\end{equation}
where $A^c$ is the complementary region to $A$. This will play an important role when computing the energy density of these states, given that it implies
\begin{equation}\label{eq:102}
\mathcal{E}_A(2\pi \ell)=
  {\rm Tr}\left(\rho_A(2\pi \ell)\, T_{tt}\right)=
  \bra{0_M}T_{tt}\ket{0_M}=0\ .
\end{equation}
There will be two distinct regimes, $\beta$ smaller or larger than $2\pi \ell$, where the energy density becomes positive or negative. The inverse temperature $\beta$ will allows us to interpolate between positive and negative energy density. In order to compute these quantities explicitly, we employ several tools previously developed in Refs. \cite{Unruh:1976db,Bisognano:1976za,Candelas:1978gf,Hislop:1981uh,Haag:1992hx,Emparan:1999gf,Casini:2011kv}.

\subsection{Half-space region}


Consider a $d$-dimensional CFT in Minkowski space-time $ds^2=-dt^2+dx^2+d\vec{y}.d\vec{y}$ and take the region $A$ at $t=0$ as the half-space $\left\lbrace (x,\vec{y}\,)\in \mathbb{R}\times \mathbb{R}^{d-2}: x>x_0 \right\rbrace$ with its boundary at $x_0$. Its causal domain is the right Rindler wedge, given by
\begin{equation}\label{eq:98}
\mathcal{D}_{HS}=
  \left\lbrace
  (t,x,\vec{y})\in \mathbb{R}\times\mathbb{R}\times\mathbb{R}^{d-2}:
  u_\pm=x\pm t>x_0
  \right\rbrace\ .
\end{equation}
To construct the operator $K_s$ in Eq. (\ref{eq:86}) we must find a well defined timelike evolution in $\mathcal{D}_{HS}$.\footnote{The Minkowski time $t$ does not give a well defined time evolution in $\mathcal{D}_{HS}$ since its range depends on the spatial coordinates, \textit{i.e.} $t\in (-x,x)$ for $x_0=0$. A well defined time evolution must be given by a real parameter $s$ subject to no spatial dependent restrictions.} This can be easily done by taking standard Rindler coordinates $(\eta,\xi)$ given by
\begin{equation}\label{eq:94}
u_\pm(\eta,\xi)=x_0+\xi\, e^{\pm \eta/\alpha}\ ,
\end{equation}
which automatically verify the constraints in (\ref{eq:98}) as long as $\xi>0$. The dimensionful constant $\alpha$ plays no fundamental role and is introduced so that $\eta$ has units of time. Since the coordinate $\eta$ gives a well defined time evolution in $\mathcal{D}_{HS}$, we can identify it with the $s$ parameter in Eq. (\ref{eq:86}) and explicitly write the operator $K_{\eta}$ generating $\eta$ translations, and its associated thermal state
\begin{equation}\label{eq:112}
K_{\eta}=\frac{1}{\alpha}
  \int d\vec{y}\int_{x_0}^{+\infty}dx
  \,(x-x_0)\,T_{tt}(0,x,\vec{y}\,)\ ,
  \qquad \qquad
  \rho_{HS}(\beta)=\frac{1}{Z}\exp\left(
  -\beta K_{\eta}
  \right) \ .
\end{equation}
The operator $K_{\eta}$ is proportional to the boost operator in the $x$ direction. For $\beta=2\pi \alpha$ the well known Unruh effect \cite{Unruh:1976db,Bisognano:1976za} implies that this thermal state becomes equivalent to the Minkowski vacuum $\ket{0_M}$ reduced to $\mathcal{D}_{HS}$, as given in Eq. (\ref{eq:85}) after replacing $\ell\rightarrow \alpha$.

We now compute the energy density of $\rho_{HS}(\beta)$
\begin{equation}\label{eq:99}
\mathcal{E}_{HS}(\beta)={\rm Tr}\left(
  \rho_{HS}(\beta)\,T_{tt}
  \right)\ ,
\end{equation}
which from Eq. (\ref{eq:102}) must vanish for $\beta=2\pi\alpha$. We first apply a conformal transformation given by the change of coordinates in Eq. (\ref{eq:94}), so that the Minkowski metric becomes
\begin{equation}\label{eq:95}
ds^2=-dt^2+dx^2+d\vec{y}.d\vec{y}=
  \left(\xi/\alpha\right)^2
  \left(
  -d\eta^2+\alpha^2dH_{d-1}^2
  \right)\ ,
\end{equation}
where $dH_{d-1}$ is the line element of a unit hyperbolic plane 
\begin{equation}\label{eq:161}
dH_{d-1}^2=
  \frac{d\xi^2+d\vec{y}.d\vec{y}}{\xi^2}\ .
\end{equation}
Applying a Weyl rescaling which removes the conformal factor in Eq. (\ref{eq:95}), the Rindler region $\mathcal{D}_{HS}$ is mapped to the entire hyperbolic space-time $\mathbb{R}\times \mathbb{H}^{d-1}$. This transformation is implemented on the Hilbert space by the unitary operator $U:\mathcal{H}_{HS}\rightarrow \bar{\mathcal{H}}_{\rm Hyp}$, where $\bar{\mathcal{H}}_{\rm Hyp}$ is the Hilbert space of the CFT in the hyperbolic space-time. We will add bar over quantities defined after the conformal transformation. 

To see how the energy density (\ref{eq:99}) transforms, we must look at the transformation of the stress tensor, that is given by\footnote{The first factor is given by the Weyl rescaling and can be easily obtained by seeing how $T_{\mu \nu}=-\frac{2}{\sqrt{-g}}\frac{\delta I_{CFT}}{\delta g^{\mu \nu}}$ transforms.}
\begin{equation}\label{eq:96}
T_{\mu \nu}=
  (\alpha/\xi)^{d-2}
  \frac{\partial X^a}{\partial x^\mu}
  \frac{\partial X^b}{\partial x^\nu}
  \left(
  U^\dagger\bar{T}_{ab}U
  -S_{ab}
  \right)
  \ ,
\end{equation}
where $x^\mu=(t,x,\vec{y})$ and $X^a=(\eta,\xi,\vec{y})$. The operator $S_{ab}$ is the anomalous contribution proportional to the identity operator which generalizes the two-dimensional Schwartzian derivative \cite{DiFrancesco:1997nk} and is different from zero only for $d$ even. Though its general expression is not known,\footnote{See however Refs. \cite{Herzog:2013ed,Brown:1977sj,Herzog:2015ioa}.} if we take the expectation value of Eq. (\ref{eq:96}) with respect to $\rho_{HS}(2\pi \alpha)$, the left-hand side vanishes due to Eq. (\ref{eq:102}) and we find
\begin{equation}\label{eq:97}
S_{ab}=
  \begin{cases} 
  \,\,{\rm Tr}\left(
  \bar{\rho}(2\pi \alpha)\,\bar{T}_{ab}
  \right) \ ,
  &\,\, {\rm for\,\,}d{\,\,\rm even} \\
  \,\,\qquad \quad 0\,\, \qquad \quad\ ,
 & \,\,\,\,\, {\rm for\,\,}d{\,\,\rm odd}\ , \\
 \end{cases}
\end{equation}
where $\bar{\rho}(\beta)=U\rho_{HS}(\beta)U^\dagger$ is the transformed state, an ordinary thermal state in the hyperbolic background $\mathbb{R}\times \mathbb{H}^{d-1}$. Since the hyperbolic space is maximally symmetric and $\bar{\rho}(\beta)$ is an equilibrium state, the thermal expectation value of $\bar{T}^a_{\,\,\,b}$ for arbitrary $\beta$ is given by
\begin{equation}\label{eq:100}
{\rm Tr}
  \left(
  \bar{\rho}(\beta)\,
  \bar{T}^a_{\,\,\,b}
  \right)=
  \frac{\bar{\mathcal{E}}(\beta)}{d-1}
  {\rm diag}\left(1-d,1,\dots,1\right)\ ,
\end{equation}
where $\bar{\mathcal{E}}(\beta)$ is the thermal energy density in the hyperbolic space-time. In this expression we have already imposed the zero trace condition of the stress tensor, which follows from the fact that the hyperbolic space-time is conformally flat and the Euler density of $\mathbb{R}\times \mathbb{H}^{d-1}$ vanishes.\footnote{Though this is true for arbitrary $d$, it is straightforward to check for the first few even values.} Using Eqs. (\ref{eq:96}-\ref{eq:100}) in Eq. (\ref{eq:99}), we obtain the following expression for the energy density of $\rho_{HS}(\beta)$
\begin{equation}\label{eq:101}
\mathcal{E}_{HS}(\beta)=
  \frac{
  \bar{\mathcal{E}}(\beta)-
  \bar{\mathcal{E}}(2\pi \alpha)}
  {\Omega(u_\pm)^d}
  \left[1+
  \frac{d}{d-1}
  \left(
  \frac{u_+-u_-}{2 \alpha\, \Omega(u_\pm)}
  \right)^2
  \right]\ ,
\end{equation}
where 
$$\Omega(u_\pm)=\left[
  \frac{(u_+-x_0)(u_--x_0)}{\alpha^2}
  \right]^{1/2}\ ,$$ 
is the conformal factor in $u_\pm$ coordinates. Since the term $\bar{\mathcal{E}}(2\pi \alpha)$ comes from Eq. (\ref{eq:97}) it vanishes for odd $d$.

As discussed in the introduction, the energy density becomes singular at the boundary of $\mathcal{D}_{HS}$ given by $u_\pm=x_0$ (\ref{eq:98}). Moreover, the sign of $\mathcal{E}_{HS}(\beta)$ is completely determined by the sign of $\bar{\mathcal{E}}(\beta)-\bar{\mathcal{E}}(2\pi \alpha)$. Since the energy density of a thermal state always increases with its temperature and ${\mathcal{E}_{HS}(2\pi \alpha)=0}$, we conclude that $\rho_{HS}(\beta)$ has negative energy density for ${\beta>2\pi \alpha}$ and positive for ${\beta<2\pi \alpha}$. All that remains to obtain an explicit expression for $\mathcal{E}_{HS}(\beta)$ is compute $\bar{\mathcal{E}}(\beta)$. To do so, we must consider a particular space-time dimension and CFT.

\addtocontents{toc}
{\protect\setcounter{tocdepth}{1}}
\subsubsection{Two-dimensional CFT}
\addtocontents{toc}
{\protect\setcounter{tocdepth}{2}}

For any two-dimensional CFT the calculation of $\bar{\mathcal{E}}(\beta)$ becomes very simple due to the fact that the hyperbolic plane in Eq. (\ref{eq:161}) becomes the real line, \textit{i.e.} $\alpha^2 dH_1^2=\left(\alpha \,d\xi/\xi\right)^2=\left[d(\alpha\ln(\xi/\alpha))\right]^2$. This means that $\bar{\mathcal{E}}(\beta)$ is the energy density of a thermal state in two-dimensional Minkwoski. The only dimensionful quantity it can depend on is $\beta$, meaning that dimensional analysis implies $\bar{\mathcal{E}}(\beta)\propto \beta^{-2}$. The proportionality constant can be determined from a standard computation that involves going to Euclidean time and compactifying on a circle of length $\beta$, which gives \cite{DiFrancesco:1997nk,Herzog:2015ioa}
\begin{equation}\label{eq:105}
\bar{\mathcal{E}}(\beta)=\frac{c\pi}{6\beta^2}\ ,
\end{equation}
where $c$ is the Virasoro central charge of the CFT. Using this in Eq. (\ref{eq:101}) we can write $\mathcal{E}_{HS}(\beta)$ in terms of the dimensionless temperature $T=2\pi\alpha/\beta$ as
\begin{equation}\label{eq:113}
\mathcal{E}_{HS}(\beta)=
  \frac{c(T^2-1)}{48\pi}
  \left[
  \frac{1}{(u_+-x_0)^{2}}+
  \frac{1}{(u_--x_0)^{2}}
  \right]\ .
\end{equation}
For any value of $T<1$ the energy density is negative.

\subsubsection{Arbitrary dimensional CFTs}


For $d=2$ the thermal energy density $\bar{\mathcal{E}}(\beta)$ was practically fixed by dimensional analysis since the hyperbolic plane $\alpha^2 dH_{d-1}^2$ is the real line. This is not the case for $d>2$, meaning that $\bar{\mathcal{E}}(\beta)$ will not only depend on $\beta$, but also on the radius of the hyperbolic plane $\alpha$. In the following we will consider particular cases of free and strongly coupled CFTs where $\bar{\mathcal{E}}(\beta)$ can be obtained explicitly.

\subsubsection*{Free theories}

For free theories we can directly quantize the classical field in the hyperbolic background and compute the thermal energy density $\bar{\mathcal{E}}(\beta)$. Let us consider a free massless scalar in $d$-dimensional Minkowski space-time, which aquires a conformal coupling to the background geometry when applying the conformal transformation (see Eq. (\ref{eq:27}) for the resulting action of the scalar field after the conformal transformation). In Appendix \ref{app:hyp} we apply standard canonical quantization to compute the thermal energy density $\bar{\mathcal{E}}(\beta)$.\footnote{The systematic canonical quantization of a thermal scalar field conformally coupled to $\mathbb{R}\times \mathbb{H}^{d-1}$ has not been presented in the literature, though substancial work has been done in Refs. \cite{Bunch:1978ka,Candelas:1978gf,Denardo:1981xa,Brown:1982hb,Pfautsch:1982hv,Page:1982fm,Bytsenko:1994bc,Moretti:1995fa,Iellici:1997yh,Haba:2007ay,Cho:2014ira,Klebanov:2011uf} . We explicitly do so in App. \ref{app:hyp} for arbitrary temperature and space-time dimensions, and compute the thermal two-point functions, energy density and partition function.} 

\begin{figure}[h]\centering
\includegraphics[scale=0.55]{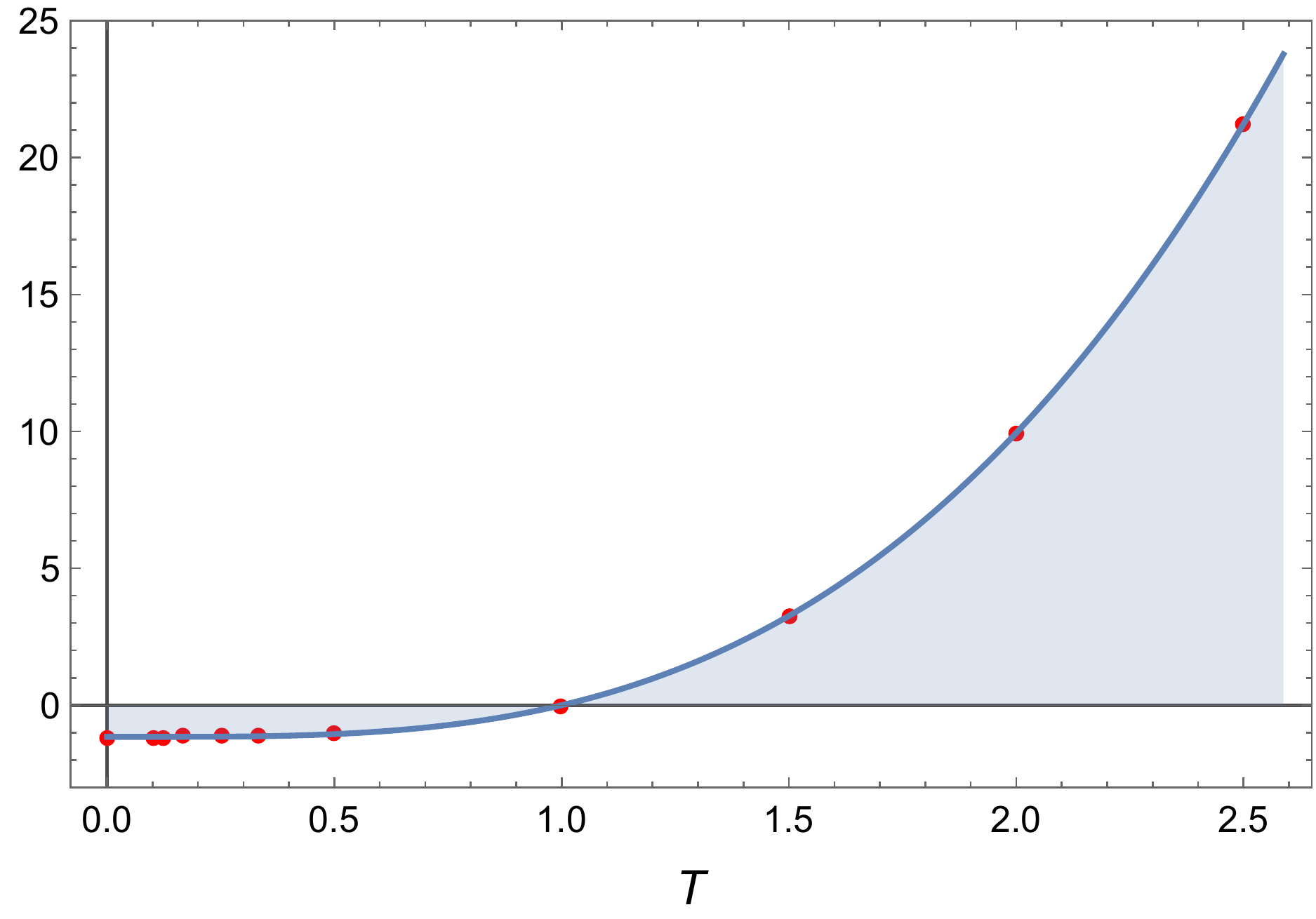}
\caption{Energy density of a scalar field conformally coupled to $\mathbb{R}\times \mathbb{H}^{d-1}$ for $d=3$ and $\alpha=1$ found from numerical integration of Eq. (\ref{eq:48}), as a function of $T=2\pi \alpha/\beta$. The red points correspond to rational values of $T$ where the integral can be solved exactly.}\label{fig:3}
\end{figure}

For the first few even values of $d>2$, we obtain the results in Table \ref{table:1}, written in terms of the dimensionless temperature $T=2\pi \alpha/\beta$. Using this in Eq. (\ref{eq:101}) we can explicitly write the energy density of $\rho_{HS}(\beta)$.
\begin{table}[]\setlength{\tabcolsep}{10pt}
\centering
\begin{tabular}{|Sc|Sc|Sc|Sc|Sc|}
\hline 
$d$ & 4 & 6 & 8  \\  \hline
$\bar{\mathcal{E}}(\beta)-\bar{\mathcal{E}}(2\pi \alpha)$  &
 $\displaystyle\frac{T^4-1}
 {480\pi^2\alpha^4}$  &
  $\displaystyle\frac{10T^6+21T^4-31}
  {60480 \pi^3\alpha^6}$  &
  $\displaystyle\frac{21T^8+100T^6+168T^4-289}
  {1209600\pi^4\alpha^8}$ 
  \\ \hline
\end{tabular}
\caption{Thermal energy density of a scalar field conformally coupled to the hyperbolic background $\mathbb{R} \times \mathbb{H}^{d-1}$, written in terms of the dimensionless temperature $T=2\pi \alpha/\beta$. As required, all the expressions vanish for $T=1$.}\label{table:1}
\end{table}
For odd $d$ the calculation of $\bar{\mathcal{E}}(\beta)$ is technically more challenging. For $d=3$, it can be written in terms of the following integral
\begin{equation}\label{eq:48}
\bar{\mathcal{E}}(\beta)=
  \frac{1}
  {4\pi^2\alpha^3\sqrt{2}}
  \int_{0}^\infty
  \frac{dv}{\left(\cosh(v)-1\right)^{5/2}}
  \left[
  \sinh(v)-
  \left(\frac{T\sinh(v/2)}{\sinh(Tv/2)}\right)^4
  \frac{\sinh(Tv)}{T}
  \right]\  ,
\end{equation}
which vanishes when $\beta=2\pi\alpha$ since there is no trace anomaly in odd dimensions (\ref{eq:97}). For several rational values of $T$ the integral can be solved exactly, the most interesting being the zero temperature case where we find 
\begin{equation}\label{eq:154}
\bar{\mathcal{E}}(\beta\rightarrow \infty)=
  -\frac{3\zeta(3)}{32\pi^4\alpha^3}\ ,
\end{equation}
with $\zeta(z)$ is the Riemann zeta function. For general $T$, we can solve through numerical integration and obtain the plot in Fig. \ref{fig:3}, the red dots corresponding to values of $T$ which allow for exact integration.  Using Eqs. (\ref{eq:48}) and (\ref{eq:101}), we obtain an integral expression for the energy density of $\rho_{HS}(\beta)$.

\subsubsection*{Strongly coupled theories}


We now move to the opposite side of the spectrum and consider strongly coupled CFTs. To do so, we restrict to theories with large number of degrees of freedom and apply the standard AdS/CFT dictionary \cite{Maldacena:1997re,Witten:1998qj,Gubser:1998bc}. In this limit, the thermal state $\bar{\rho}(\beta)$ in $\mathbb{R}\times \mathbb{H}^{d-1}$ will be dual to a black hole with a hyperbolic horizon. The energy density $\bar{\mathcal{E}}(\beta)$ can be computed from a standard gravity calculation of the quasi-local stress tensor of the black hole \cite{Brown:1992br,Balasubramanian:1999re}.

For holographic CFTs dual to pure Einstein gravity we compute the quasi-local stress tensor of the appropriate hyperbolic black hole in App. \ref{app:quasi} and obtain the following result for the energy density 
\begin{equation}\label{eq:106}
\bar{\mathcal{E}}(\beta)-\bar{\mathcal{E}}(2\pi \alpha)=
  \frac{a_d^*(d-1)g(\beta)^{d-2}\left(g(\beta)^2-1\right)}
  {{\rm Vol}(S^{d-1})\alpha^d}
  \ ,
  \qquad 
  g(\beta)=
  \frac{T+\sqrt{T^2+d(d-2)}}{d}\ ,
\end{equation}
where $T=2\pi \alpha/\beta$, ${\rm Vol}(S^{d-1})=2\pi^{d/2}/\Gamma(d/2)$, and $a_d^*$ is the generalized central charge given by \cite{Nishioka:2018khk}
\begin{equation}\label{eq:120}
a_d^*=
  \begin{cases} 
  \qquad \qquad \,\,\,\,\,
  A_d
  \qquad \quad \,\,\,\, 
  \ ,
  & {\rm for\,\,d\,\,even} \vspace{6pt}\\
  \,\,(-1)^{\frac{d-1}{2}}\ln[
  Z(S^d)]/2\pi\ ,
  &  {\rm for\,\,d\,\,odd}\ , \\
 \end{cases}
\end{equation}
with $A_d$ the $A$-type trace anomaly of the stress tensor (see Ref. \cite{Myers:2010tj} for conventions) and $Z(S^d)$ the regularized vacuum partition function of the CFT placed on a unit $d$-dimensional sphere (see Ref. \cite{Pufu:2016zxm} for $S^3$ examples). Using this in Eq. (\ref{eq:101}) we can write the energy density of $\rho_{HS}(\beta)$ for any space-time dimension and temperature.
  
The same calculation can be done for holographic CFTs dual to Gauss-Bonnet gravity. In App. \ref{app:quasi} we explicitly compute $\bar{\mathcal{E}}(\beta)$ for $d=4$ and arbitrary values of $\beta$. In the zero temperature limit the expression simplifies and becomes
\begin{equation}\label{eq:126}
\bar{\mathcal{E}}(\beta\rightarrow \infty)-
  \bar{\mathcal{E}}(2\pi \alpha)=
  -\frac{3}{2\pi^2\alpha^4}
  \left(\frac{c^2}{5c-a}\right)\ ,
\end{equation}
where $a$ and $c$ are the coefficients of the terms appearing in the trace anomaly of the four dimensional CFT (not to be confused with the two-dimensional Virasoro central charge). These are constrained by the Hofman-Maldacena bounds according to \cite{Hofman:2008ar}
\begin{equation}\label{eq:160}
\frac{1}{3}\le \frac{a}{c}\le \frac{31}{18}\ ,
\end{equation}
so that right-hand side of Eq. (\ref{eq:126}) is always negative. Taking $c=a$ it is straightforward to check that it reduces to the Einstein gravity result of Eq. (\ref{eq:106}) in the zero temperature limit.

\subsection{Ball region}


A very similar construction can be applied to a ball $\mathcal{B}$ of radius $R$ at $t=0$. Its causal domain can be written as
\begin{equation}\label{eq:67}
\mathcal{D}_{\mathcal{B}}=
  \left\lbrace \,\,
  (t,r,\theta_i)\in 
  \mathbb{R}\times\mathbb{R}_{\ge 0}\times S^{d-2}
  \,\,:\,\, 
  |w_\pm|=|r\pm t|\le R\, 
  \right\rbrace\ ,
\end{equation}
where $\theta_i$ are the angles on the unit sphere $S^{d-2}$ and we are implicitly assuming ${w_++w_-\ge 0}$.\footnote{For $d=2$ this region is diamond in the $(t,x)$ plane, while for $d=3$ it is obtained from rotating this diamond along the $t$ axis.} Same as before, we must find a well defined timelike coordinate in $\mathcal{D}_{\mathcal{B}}$ to construct the operator $K_s$ in Eq. (\ref{eq:86}). The constraint in (\ref{eq:67}) is automatically verified if we define the coordinates $(\tau,u)$ according to
\begin{equation}\label{eq:103}
w_{\pm}(\tau,u)=R\tanh\left(\frac{Ru\pm \tau}{2R}\right)\ ,
\end{equation}
where $r\ge 0$ implies $u\ge 0$.\footnote{This is the same change of coordinates considered in Ref. \cite{Casini:2011kv} but written in a more compact way.} The timelike coordinate $\tau$ is unconstrained and gives a well defined time evolution in $\mathcal{D}_{\mathcal{B}}$. We can identify $\tau$ with the $s$ parameter in Eq. (\ref{eq:86}) and explicitly write the operator $K_{\tau}$ generating $\tau$ translations and its associated thermal state
\begin{equation}\label{eq:117}
K_{\tau}=
  \int_{S^{d-2}}d\theta_i
  \int_0^Rdr\,r^{d-2}\left(
  \frac{R^2-r^2}{2R^2}
  \right)T_{tt}(0,r,\theta_i)\ ,
  \quad \quad \,\,\,\,
  \rho_{\mathcal{B}}(\beta)=\frac{1}{Z}\exp\left(
  -\beta K_{\tau}
  \right) \ .
\end{equation}
For $\beta=2\pi R$ this thermal state is equivalent to the Minkowski vacuum $\ket{0_M}$ reduced to $\mathcal{D}_{\mathcal{B}}$, as given in Eq. (\ref{eq:85}) after replacing $\ell\rightarrow R$ \cite{Hislop:1981uh,Haag:1992hx,Casini:2011kv}.

To compute the energy density of $\rho_{\mathcal{B}}(\beta)$
\begin{equation}\label{eq:104}
\mathcal{E}_{\mathcal{B}}(\beta)={\rm Tr}\left(
  \rho_{\mathcal{B}}(\beta)\,
  T_{tt}
  \right)\ ,
\end{equation}
we apply a conformal transformation given by the change of coordinates in Eq. (\ref{eq:103}), so that the Minkowski metric becomes
\begin{equation}\label{eq:152}
ds^2=-dt^2+dr^2+r^2ds^2_{S^{d-2}}=
  \frac{-d\tau^2+R^2dH_{d-1}^2}
  {\left(\cosh(\tau/R)+\cosh(u)\right)^2}\ ,
\end{equation}
where $dH_{d-1}$ is the line element of a unit hyperbolic plane in a different set of coordinates
\begin{equation}\label{eq:163}
dH_{d-1}^2=du^2+\sinh^2(u)ds^2_{S^{d-2}}\ .
\end{equation}
Applying a Weyl rescaling which removes the conformal factor, the region $\mathcal{D}_{\mathcal{B}}$ is mapped to the entire hyperbolic space-time $\mathbb{R}\times \mathbb{H}^{d-1}$.

Same as before, we will compute the energy density of $\rho_\mathcal{B}(\beta)$ by first looking at the transformation of the stress tensor, that will have an analogous expression to Eq. (\ref{eq:96})
$$T_{\mu \nu}=
  (\cosh(\tau/R)+\cosh(u))^{d-2}
  \frac{\partial X^a}{\partial x^\mu}
  \frac{\partial X^b}{\partial x^\nu}
  \left(
  U^\dagger\bar{T}_{ab}U
  -S_{ab}
  \right)
  \ ,$$
where now $x^\mu=(t,r,\theta_i)$ and $X^a=(\tau,u,\theta_i)$. Using the same argument as before, the anomalous terms $S_{ab}$ is given by Eq. (\ref{eq:97}) replacing $\alpha\rightarrow R$. Using this and the thermal stress tensor in Eq. (\ref{eq:100}), we can use the change of coordinates in Eq. (\ref{eq:103}) and write the energy density of $\rho_{\mathcal{B}}(\beta)$ as
\begin{equation}\label{eq:109}
\mathcal{E}_{\mathcal{B}}(\beta)=
  \frac{\bar{\mathcal{E}}(\beta)-
  \bar{\mathcal{E}}(2\pi R)}
  {\Omega(w_\pm)^{d}}
  \left[
  1+\frac{d}{d-1}
  \left(\frac{w_+^2-w_-^2}{4R^2\,\Omega(w_\pm)}\right)^2
  \right]\ ,
\end{equation}
where
$$\Omega(w_\pm)=\left[
  \frac{(R^2-w_+^2)(R^2-w_-^2)}{4R^4}
  \right]^{1/2} \ , $$
is the conformal factor in $w_\pm$ coordinates. The energy density is again divergent at the boundary of $\mathcal{D}_{\mathcal{B}}$, given by $|w_\pm|=R$ (\ref{eq:67}). Its sign is determined by $\bar{\mathcal{E}}(\beta)-\bar{\mathcal{E}}(2\pi R)$, meaning that the energy will be negative for ${\beta>2\pi R} $ and positive otherwise. We can obtain explicit expressions for $\mathcal{E}_\mathcal{B}(\beta)$ using the results of Eqs. (\ref{eq:48}), (\ref{eq:106}), (\ref{eq:126}) and Table \ref{table:1}, making the replacement $\alpha \rightarrow R$. 

For a two-dimensional CFT the ball becomes a segment of length $2R$. The radial coordinate gets replaced by a cartesian coordinate $r\rightarrow x\in \mathbb{R}$, meaning that $u$ can now take any real value. The energy density of $\rho_{\mathcal{B}}(\beta)$ can be explicitly written from Eqs. (\ref{eq:105}) and (\ref{eq:109}) in terms of $T=2\pi R/\beta$ as
\begin{equation}\label{eq:118}
\mathcal{E}_{\mathcal{B}}(\beta)=
  \frac{c(T^2-1)}{48\pi R^2}
  \left[
  \left(\frac{2R^2}{R^2-w_+^2}\right)^2+
  \left(\frac{2R^2}{R^2-w_-^2}\right)^2
  \right]\ ,
\end{equation}
which is explicitly negative when $T<1$.

\section{Energy measured by observer}
\label{sec:Obs}

In the previous section we have constructed localized states defined in the Hilbert space $\mathcal{H}_A$, where $A$ is either the half-space or a ball in Minkowski at $t=0$. A disturbing feature of these states is that their energy density diverges at the boundary of $\mathcal{D}_A$, which raises the question of whether they are physically allowed. To assess this, we consider an observer restricted to $\mathcal{D}_A$ and compute the energy measured along a given trajectory, which must be finite and in agreement with the Quantum Energy Inequalities (QEIs) present in the literature.

For simplicity we take $A$ as a ball $\mathcal{B}$ and consider a static observer fixed at the center $r=0$. A similar analysis follows for the half-space and observers moving along constant speed trajectories. The measured energy is given by 
\begin{equation}\label{eq:107}
E_{\rm obs}\left[\rho,\varphi(t)\right]=
  \int_\mathbb{R} dt\,\varphi(t)\,
  {\rm Tr}\big(
  \rho\, T_{tt}(t,r=0)
  \big)
  \ ,
\end{equation}
where $\varphi(t)$ is the weight function characterizing the measurement apparatus. Evaluating for the energy density of $\rho_{\mathcal{B}}(\beta)$ in Eq. (\ref{eq:109}), we find
\begin{equation}\label{eq:110}
E_{\rm obs}\left[\rho_\mathcal{B},\varphi(t)\right]=
  \left[\bar{\mathcal{E}}(\beta)-
  \bar{\mathcal{E}}(2\pi R)\right]
  \int_{\mathbb{R}} dt\,\varphi(t)\,
  \left(\frac{2R^2}{R^2-t^2}\right)^{d}\ .
\end{equation}
Given that the state $\rho_{\mathcal{B}}(\beta)$ is only defined in $\mathcal{D}_B$, this expression is only valid if the weight function is compactly supported in $|t|<R$. The simplest choice for $\varphi(t)$ is to consider the characteristic function of the interval of $|t|<R$, such that every point along the trajectory in $\mathcal{D}_\mathcal{B}$ is assigned equal weight. However, this naive choice of $\varphi(t)$ makes the measured energy in Eq. (\ref{eq:110}) divergent. In particular, for $\beta>2\pi R$ it becomes minus infinity in direct contradiction with numerous QEIs in the literature. 

Similar scenarios have been previously recognized and addressed in the literature, despite generating some initial confusion.\footnote{See the comments of Ref. \cite{Fewster:2004gs} with respect to the claims in Ref. \cite{Krasnikov:2004sj}.} Note that the conflict is resolved if we require the weight function to be smooth ($\varphi(t)\in C^\infty(\mathbb{R})$). Since it must also be compactly supported in $\mathcal{D}_\mathcal{B}$, this means it approaches the boundary $|t|=R$ faster than any polynomial and expression (\ref{eq:110}) yields a finite value. This smoothness constraint on $\varphi(t)$ is not a capricious choice, but can be understood in several ways.

The physical argument in favor of the smoothness requirement of $\varphi(t)$ comes from the fact that this function is determined by the measurement apparatus used to detect energy. Any real device will not have a discontinuous profile given that there is always a relaxation time associated to its variations. Moreover $T_{tt}(t,0)$ is an operator valued distribution whose domain is given by suitable smooth test functions. It is therefore no surprise that some states yield incoherent expectation values when $T_{tt}(t,0)$ is evaluated on functions that are outside its domain. Finally, all the rigorous QEIs derived in the literature require the weight function to be smooth. If one tries to relax this requirement, it can be shown that the bounds fails to be true (\textit{e.g.} see Sec. 4.2.4 of Ref. \cite{Fewster:2004nj}) and the energy measured by an observer is unbounded from below.

Having established the smoothness requirement of $\varphi(t)$, the question remains whether the amount of negative energy measured by the observer when $\beta>2\pi R$ is consistent with the QEIs present in the literature. We will consider this question for arbitrary two-dimensional CFTs and a massless scalar field in arbitrary dimensions, where QEIs are available.

\subsubsection*{Two-dimensional CFTs}

Consider the following QEI valid for any two-dimensional CFT
\begin{equation}\label{eq:147}
E_{\rm obs}\left[\rho,\varphi(t)\right]
  \ge -\frac{c}{6\pi}
  \int_{\mathbb{R}}
  dt\left(\frac{d}{dt}\sqrt{\varphi(t)}\right)^2\ ,
\end{equation}
where $\varphi(t)$ must be an even function of the Schwartz type, \textit{i.e.} smooth and all its derivatives must decay at infinity faster than any polynomial. Notice that only the left hand side depends on the state under consideration. This bound was rigorously proven in Ref. \cite{Fewster:2004nj} where it was also shown to be sharp, \textit{i.e.} for a fixed weight function $\varphi(t)$ there is always a state which saturates the inequality. An alternative derivation, which also includes a correction term for mixed states, was given in Ref. \cite{Blanco:2017akw} from the monotonicity property of relative entropy.\footnote{See Ref. \cite{Levine:2016bpj} for another derivation valid for holographic CFTs.}

Evaluating the inequality for the energy density of $\rho_\mathcal{B}(\beta)$ (\ref{eq:118}) in the zero temperature limit (where it has its most negative value) and redefining the weight function as ${h(t)=\sqrt{\varphi(t)}}$ we can write inequality (\ref{eq:147}) as
\begin{equation}\label{eq:119}
\int_{-1}^1dv\,
  h(vR)
  \left[
  -\frac{d^2}{dv^2}
  -\frac{1}{(1-v^2)^2}
  \right]h(vR)
  \ge 0\ ,
\end{equation}
where we have used that $h(t)$ must be complactly supported in $|t|<R$ and changed variables to $v=t/R$. Following Ref. \cite{Fewster:1999kr} we can view this inequality as a statement of the non-negativity of the operator in Eq. (\ref{eq:119}) acting on the Hilbert space of smooth functions on the interval $v\in(-1,1)$. If we rewrite this operator in the following way
\begin{equation}\label{eq:134}
B=\frac{d}{dv}+\frac{1-v\,{\rm arctanh}(v)}
  {(v^2-1){\rm arctanh}(v)}
  \qquad \Longrightarrow \qquad
  \left[
  -\frac{d^2}{dv^2}
  -\frac{1}{(1-v^2)^2}
  \right]=B^\dagger B\ge 0\ ,
\end{equation}
we explicitly see that the left-hand side of Eq. (\ref{eq:119}) is positive. We conclude that the QEI (\ref{eq:147}) holds for the energy density of $\rho_{\mathcal{B}}(\beta)$ for any temperature and smooth weight function.

It is natural to consider what happens to the above argument if we add a little bit more of negative energy. Will the QEI (\ref{eq:147}) be violated? To consider this, we multiply the energy density of $\rho_{\mathcal{B}}(\beta\rightarrow \infty)$ with a positive factor $\lambda>1$, \textit{i.e.} $\mathcal{E}_{\mathcal{B}}\rightarrow \lambda \,\mathcal{E}_\mathcal{B}$. In this case, the QEI can be written as
\begin{equation}\label{eq:133}
\int_{-1}^1dv\,
  h(vR)
  \left[
  -\frac{d^2}{dv^2}
  -\frac{\lambda}{(1-v^2)^2}
  \right]h(vR)
  \ge 0\ .
\end{equation}
Let us write the most general differential operator $B(\lambda)$ given by
$$B(\lambda)=\frac{d}{dv}+U(v)\ ,
  \qquad {\rm such\,\,that}\qquad
  B^\dagger(\lambda) B(\lambda)=
  -\frac{d^2}{dv^2}
  -\frac{\lambda}{(1-v^2)^2}\ .$$
From this constraint is straightforward to show that the function $U(v)$ must be real and satisfy the following differential equation
$$
  U'(v)=U^2(v)+\frac{\lambda }{(1-v^2)^2}
  \ .$$
Solving this differential equation one finds that the solution is real only if $\lambda\le 1$, where for $\lambda=1$ the operator becomes the one in Eq. (\ref{eq:134}). This allows us to conclude that the energy density of $\rho_\mathcal{B}(\beta)$ in the zero temperature limit is the maximum amount allowed by the CFT.

\subsubsection*{Massless scalar field}

For a massless scalar field a QEI for arbitrary space-time dimensions was derived in Ref. \cite{Fewster:1998pu}, that simplifies for even values and can be written as
\begin{equation}\label{eq:111}
E_{\rm obs}\left[\rho,\varphi(t)\right]
  \ge -
  \frac{{\rm Vol}(S^{d-2})}
  {2d\,(2\pi)^{d-1}}
  \int_{\mathbb{R}}dt
  \left(
  \frac{d^{d/2}}{dt^{d/2}}\sqrt{\varphi(t)}
  \right)^2\ ,
\end{equation}
where $\varphi(t)$ must be even and of the Schwartz type. In contrast to the two-dimensional case, this inequality is not expected to be sharp. Trying to use the same methods to check the validity of the QEI (\ref{eq:111}) becomes complicated due to the higher order derivatives. Instead we can consider a particular set of weight function $\varphi(t)$, that we take to be zero for $|t|\ge R$ and have the following values when $|t|<R$
\begin{equation}\label{eq:65}
\begin{aligned}
\varphi_1(v)&=
  \exp\left(\frac{1}{v^2-1}\right)\ ,
\qquad  \quad  \,\,\,
\varphi_2(v)=
  \left[
  1+\exp\left(\frac{2v^2-1}{v^2(1-v^2)}\right)
  \right]^{-1}\ ,\\ \\[0.03ex]
\varphi_3(v)&=
  \left[I_0\left(\frac{1}{1-v^2}\right)
  \right]^{-1}\ ,
\qquad 
\varphi_4(v)=
  \frac{(v^2+1)\exp\left[4v/(v^2-1)\right]}
  {\left(
  (v^2-1)
  \left(1+\exp\big[4v/(v^2-1)\big]\right)
  \right)^2}\ ,
\end{aligned}
\end{equation}
where $v=t/R$ and $I_0(z)$ is the modified Bessel function of the first kind. All of these functions are smooth and by definition have compact support in $\mathcal{D}_\mathcal{B}$. 

Evaluating the left-hand side using Eq. (\ref{eq:110}) for these weight functions, both integrals appearing in the QEI (\ref{eq:111}) can be written independently of $R$ and solved numerically. Apart from considering the functions $\varphi_i(t)$ in Eqs. (\ref{eq:65}) we have raised them to positive powers and multiplied by even positive polynomials. The most constraining bounds we where able to obtain from the QEI for the factor $\bar{\mathcal{E}}(\beta)-\bar{\mathcal{E}}(2\pi R)$ in Eq. (\ref{eq:110}) are given in Table \ref{table:3}.
\begin{table}[h] \setlength{\tabcolsep}{8pt}
\centering
\begin{tabular}{|Sc|Sc|Sc|Sc|Sc|}
\hline 
$d$ & 4 & 6 & 8  \\  \hline
$\left.
 \bar{\mathcal{E}}(\beta)-\bar{\mathcal{E}}(2\pi R)
 \right|_{\rm min}$  &
  $\displaystyle
  57\left(\frac{-1}{480\pi^2R^4}\right) $  &
  $\displaystyle
  316\left(\frac{-31}{60480\pi^3R^6}\right) $ &
  $\displaystyle
  1233
  \left(\frac{-289}{1209600\pi^4R^8}\right)$
  \\ \hline
\end{tabular}
\caption{Lower bound on $\bar{\mathcal{E}}(\beta)-\bar{\mathcal{E}}(2\pi R)$ obtained from numerically evaluating the QEI (\ref{eq:111}) for the energy in Eq. (\ref{eq:110}) and the weight functions in Eqs. (\ref{eq:65}). The most constraining bound comes from taking $\left(\varphi_1(t)\right)^q$ with $q_{d=4}=2.57$, $q_{d=6}=5.2$ and $q_{d=8}=8$.}\label{table:3}
\end{table}

The factors in parenthesis in each column are the actual minimum values of  $\bar{\mathcal{E}}(\beta)-\bar{\mathcal{E}}(2\pi R)$ obtained from the transformed states $\bar{\rho}(\beta)$ (found in Table \ref{table:1} after replacing $\alpha\rightarrow R$). Since all the factors multiplying these expressions are larger than one, we find no evidence implying that the energy density of $\rho_\mathcal{B}(\beta)$ is inconsistent with the QEIs in Eq. (\ref{eq:111}). 

The conclusion of this section is that despite the singular behavior of the energy density of the states $\rho_{HS}(\beta)$ and $\rho_\mathcal{B}(\beta)$ at the boundary of the regions, they are perfectly reasonable and in agreement with the QEIs in the literature.

\section{Holographic description of localized states}
\label{sec:HoloDis}

\begin{figure}[h]\centering
\includegraphics[scale=0.80]{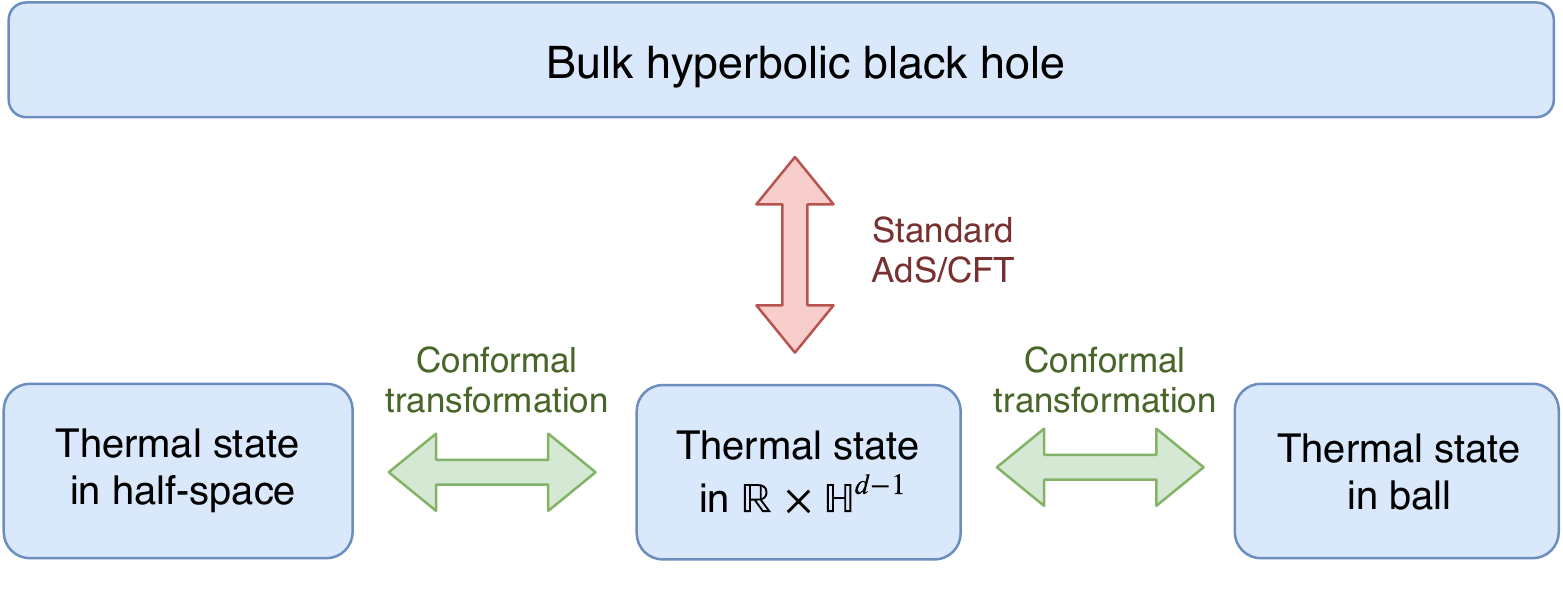}
\caption{Diagram representing the connection between the black hole with hyperbolic horizon and the localized thermal states in the half-space and a ball in Minkowski space-time. Through this series of maps the negative energy density of the localized states is mapped to the negative mass of the hyperbolic black hole.}\label{fig:5}
\end{figure}

We have focused our study on the states $\rho_{HS}(\beta)$ and $\rho_\mathcal{B}(\beta)$ that are defined on the half-space and a ball on Minkowski respectively. In both cases they can be mapped via (different) conformal transformations to an ordinary thermal state in the hyperbolic space-time $\mathbb{R}\times \mathbb{H}^{d-1}$. Applying the standard AdS/CFT dictionary this thermal state has a description in terms of a black hole with hyperbolic horizon. This means that the properties of the localized states can be mapped to the hyperbolic black hole (and vice versa) via the conformal transformation plus the standard AdS/CFT dictionary (see Fig. \ref{fig:5} for a diagram illustrating the connection).

It is particurlarly interesting to consider the mapping of the negative energy densities discussed in the previous sections. The stress tensor of the localized states is mapped to the quasi-local gravitational stress tensor \cite{Brown:1992br,Balasubramanian:1999re} of the black hole\footnote{More precisely, the energy of the localized thermal state is mapped to the quasi-local gravitational stress tensor minus its Casimir contribution.}, which will be negative for $\beta<2\pi \ell$.\footnote{The ADM mass of the hyperbolic black hole always has a range of $\beta$ for which it is negative. However, when computing it through the counter-term method, there is an addititonal positive Casimir contribution for even $d$. For $d=6,8,10,\dots$ this gives a new smaller range of $\beta$ for which the mass is negative, while for $d=2,4$ the mass is always positive, see App. \ref{app:quasi} for details.} Since the mass of the black hole is obtained by integrating this quantity it means that the negative energy density of the localized states results in a negative mass for the hyperbolic black hole. Though this negative mass has been noted long ago in the literature, its meaning has been very poorly understood. From this perspective the negative mass of the black hole is not only natural but expected, given that it arises from the negative energy density of the states in the CFT. For the zero temperature case and the ball region this connection has been recently made in Ref. \cite{Rosso:2018yax}. There it was also suggested that the relation could be generalized to arbitrary temperatures. In this work, we confirm the claims of Ref. \cite{Rosso:2018yax} and make the connection between the localized states and the black holes much more clearer and precise.

A natural question that arises is whether we can make sense of the mapping between the localized thermal states and the hyperbolic black holes without having to go through the intermediate step of the thermal state in $\mathbb{R}\times \mathbb{H}^{d-1}$. Can we map the black hole quantities directly to $\rho_{HS}(\beta)$ and $\rho_{\mathcal{B}}(\beta)$? There is a naive way of doing so. Let us see how it works for the case of the half-space. The hyperbolic black hole solution can be written as
\begin{equation}\label{eq:130}
ds^2=
  -V(\rho)(L/\alpha)^2d\eta^2+
  \frac{d\rho}{V(\rho)}+
  \rho^2\left(\frac{d\xi^2+d\vec{y}.d\vec{y}}{\xi^2}\right)\ ,
\end{equation}
where $\rho$ is the radial coordinate with the AdS boundary of radius $L$ located at $\rho\rightarrow +\infty$. The outer horizon is determined from $V(\rho_+)=0$ where the function $V(\rho)$ depends on the particular gravitational theory under consideration (see App. \ref{app:quasi} for Einstein and Gauss-Bonnet examples). Since the solution must be asymptotically AdS it must have the following behavior
$$
 {\rm For}\,\,\,\rho\gg L
  \qquad \Longrightarrow \qquad
  V(\rho)=\left(
  \frac{\rho}{L}\right)^2+\,\dots$$
The time coordinate $\eta$ has been rescaled so that we recover the hyperbolic plane of radius $\alpha$ at the boundary. Let us first review how we can take the standard boundary limit to obtain the ordinary thermal state in the hyperbolic background. Expanding the bulk metric for $\rho\gg \alpha,L$ we find
$${\rm For}\,\,\, \rho\gg \alpha,L
  \qquad \Longrightarrow \qquad
  ds^2=
  \left(\frac{\rho}{\alpha}\right)^2
  \left[
  -d\eta^2+
  \alpha^2
  \left(\frac{d\xi^2+d\vec{y}.d\vec{y}}{\xi^2}\right)
  \right]+\,\dots$$
Removing the conformal factor $(\rho/\alpha)^2$ in the leading term contribution we obtain the boundary geometry $\mathbb{R}\times \mathbb{H}^{d-1}$ where the thermal state is defined. All the black hole quantities can be mapped to the properties of the thermal state (in App. \ref{app:quasi} we map the quasi-local stress tensor in this way).

We now consider a different way of taking this limit that will enable us to recover the Rindler metric at the boundary. To do so, let us tune the coordinates $\rho$ and $\xi$ so that $\rho\gg \xi,L$. For any value of $\xi$ (which is real and positive), we can take an appropriate $\rho$ so that this is the case. The leading contribution to the bulk metric in this limit is given by
\begin{equation}\label{eq:131}
{\rm For}\,\,\, \rho\gg \xi,L
  \qquad \Longrightarrow \qquad
  ds^2=
  \left(\frac{\rho}{\xi}\right)^2
  \Big[
  -(\xi/\alpha)^2d\eta^2+
  d\xi^2+d\vec{y}.d\vec{y}\,
  \Big]+\,\dots
\end{equation}
Removing the conformal factor $(\rho/\xi)^2$ we recognize the line element of the Rindler space-time describing the causal domain of the Minkowski half space. This means that the hyperbolic black hole can be directly mapped to $\rho_{HS}(\beta)$ by taking this non-standard boundary limit, which results in the conformal factor $(\rho/\xi)^2$ instead of the more standard $(\rho/\alpha)^2$. In this way, we can avoid the CFT in the hyperbolic background and go directly to the Rindler wedge. A completely analogous discussion holds for the localized thermal state define in the ball.

\subsection{Massless hyperbolic black hole}
\label{sec:massless}

As shown in Refs. \cite{Emparan:1999gf,Casini:2011kv} the mapping between the localized thermal states and the hyperbolic black hole can be extended when the inverse temperature is given by $\beta=2\pi \ell$ with $\ell=\alpha$ or $R$. In the following, we will review and expand on the details of this construction. For these temperatures the well known Unruh effect \cite{Unruh:1976db,Bisognano:1976za} implies that the localized thermal states in Eqs. (\ref{eq:112}) and (\ref{eq:117}) are equivalent to the Minkowski vacuum reduced to each of the regions, as described in Eq. (\ref{eq:85}).

The essential bulk feature is that the hyperbolic black hole has vanishing mass\footnote{For even $d$ the black hole mass does not actually vanish when $\beta=2\pi\ell$ due to the Casimir type contribution, see App. \ref{app:quasi}. However, we will still refer to this geometry as the massless hyperbolic black hole.} and is in fact a section of pure AdS. This means that the radial function of the black hole is \textit{always} given by $V(\rho)=(\rho/L)^2-1$, for any covariant theory of gravity with negative cosmological constant \cite{Casini:2011kv}. For any other temperature different from $\beta=2\pi \ell$ the function $V(\rho)$ will have a complicated expression which depends on the gravity theory. 

\begin{figure}[h]\centering
\includegraphics[scale=0.70]{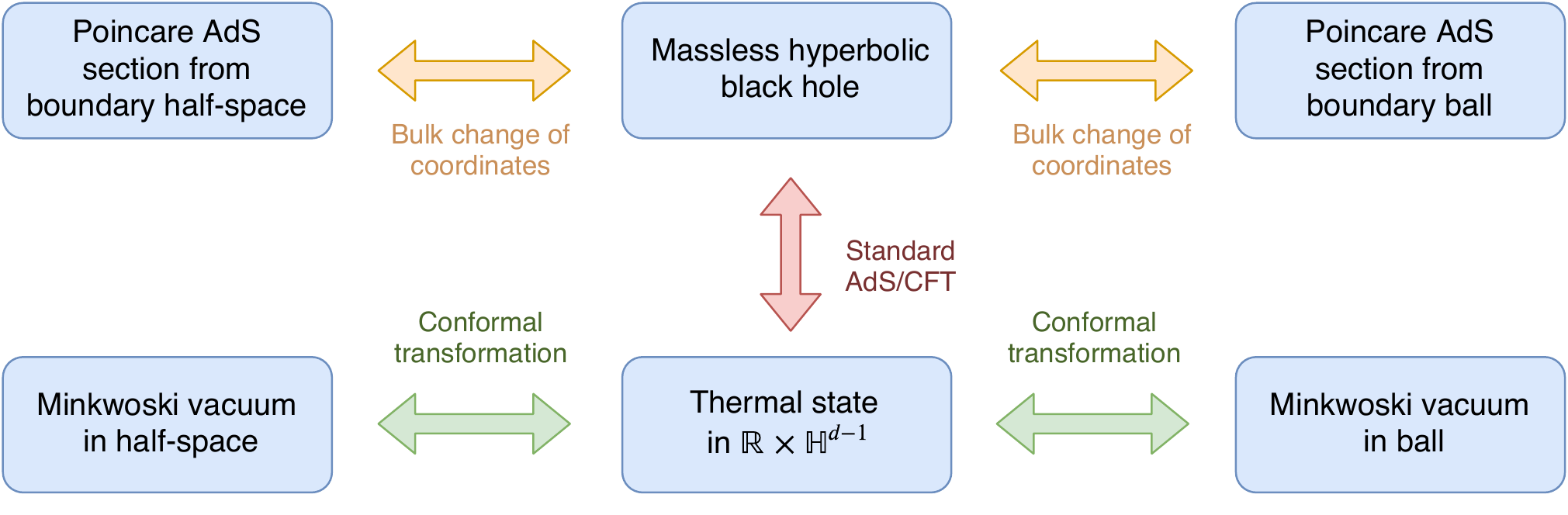}
\caption{Extension of the diagram in Fig. \ref{fig:5} for the case in which $\beta=2\pi \ell$ with $\ell=\alpha$ or $R$, where the localized thermal states become equivalent to the Minkowski vacuum reduced to the region. The hyperbolic black hole becomes massless and can be mapped to a section of Poincare AdS determined from the ball or half-space boundary regions in Minkowski.}\label{fig:6}
\end{figure}

The conformal transformations in Fig. \ref{fig:5} can be uplifted to a bulk change of coordinates so that we get the extended diagram in Fig. \ref{fig:6}. The massless hyperbolic black hole will be mapped to a section of the Poincare patch of AdS determined by the boundary half-space or ball. Let us see how this works in either case.

\subsubsection*{Half-space region}

\begin{figure}[h]
\centering
\includegraphics[height=2.50 in]{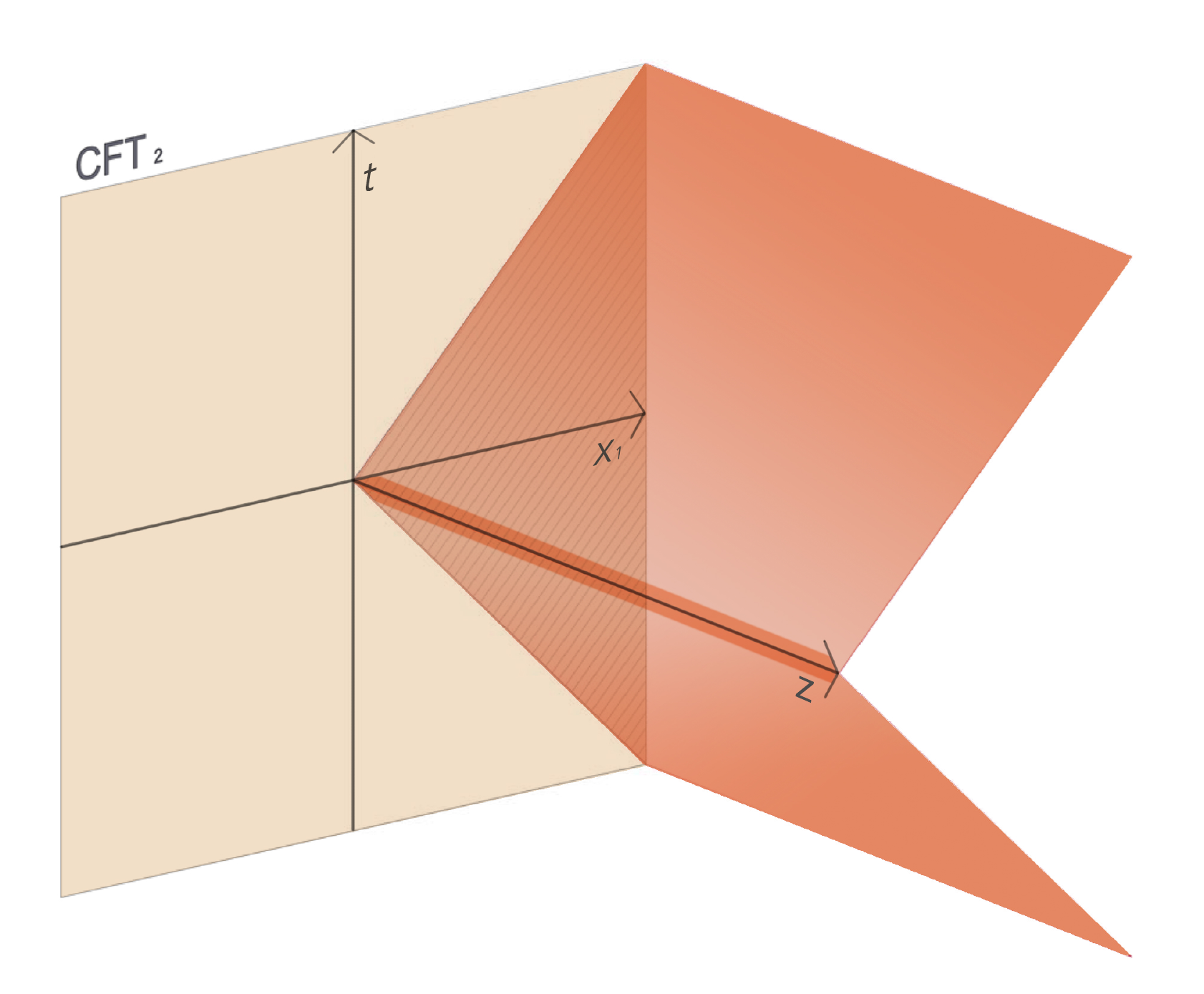}
\includegraphics[height=2.70 in]{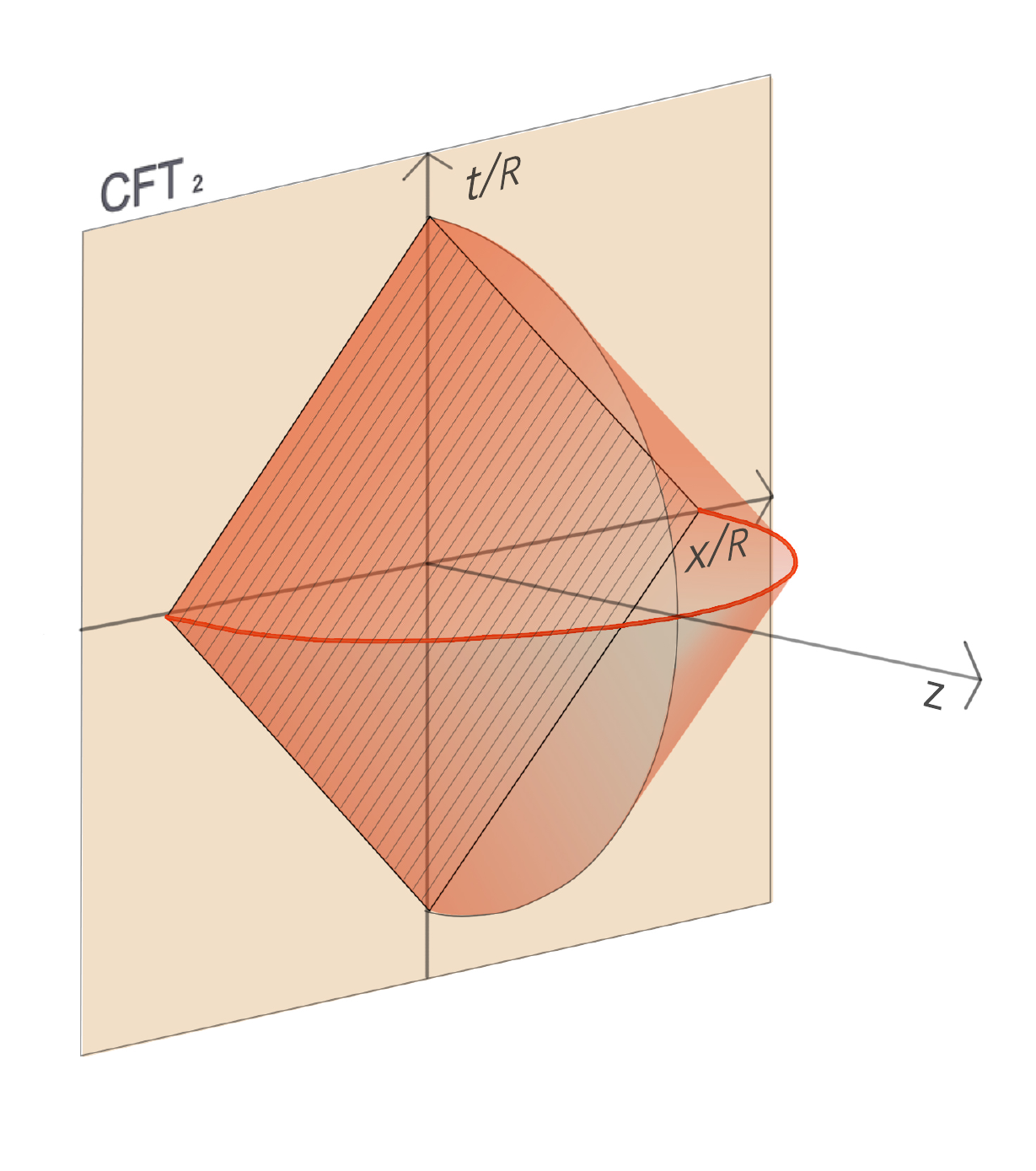}
\caption{Diagram of the horizon of the massless hyperbolic black hole in the Poincare AdS section given in Eqs. (\ref{eq:139}) and (\ref{eq:136}), after the change of coordinates in Eq. (\ref{eq:132}) (left diagram) and Eq. (\ref{eq:138}) (right diagram). Notice that for $z=0$ it intersects with the boundary of $\mathcal{D}_{HS}$ and $\mathcal{D}_{\mathcal{B}}$ respectively. In both diagrams the bulk horizon at fixed time $t=0$ has been highlighted in red.}\label{fig:4}
\end{figure}

Starting from the massless hyperbolic black hole
\begin{equation}\label{eq:143}
ds^2=-\left(\frac{\rho^2-L^2}{\alpha^2}\right)
  d\eta^2+
  \left(\frac{L^2}{\rho^2-L^2}\right)d\rho^2+
  \rho^2\left(\frac{d\xi^2+d\vec{y}.d\vec{y}}{\xi^2}\right)\ ,
\end{equation}
consider the following bulk change of coordinates \cite{Emparan:1999gf}
\begin{equation}\label{eq:132}
z=\frac{L\xi}{\rho}\ ,
  \qquad \qquad
  u_\pm=x\pm t=
  \sqrt{\frac{\rho^2-L^2}{\rho^2}}\xi e^{\pm \eta/\alpha}\  .
\end{equation}
Notice that as we take the boundary limit $\rho\rightarrow +\infty$ we recover the boundary conformal transformation in Eq. (\ref{eq:94}), meaning that this bulk change of coordinates is equivalent to the conformal transformation in the boundary theory. The inverse transformation can be written as
$$
  \rho=\frac{L }{z}\sqrt{z^2+u_+u_-}\ ,
  \qquad \qquad
  \xi=\sqrt{z^2+u_+u_-}\ ,
  \qquad \qquad
  \eta=\frac{\alpha}{2}\ln(u_+/u_-)
  \ .$$
Applying this to the massless hyperbolic black hole the metric becomes
\begin{equation}\label{eq:140}
ds^2=
  \left(\frac{L}{z}\right)^2
  \left(
  dz^2+du_+du_-+d\vec{y}.d\vec{y}\,\right)\ ,
\end{equation}
that is pure AdS in Poincare coordinates. However, the resulting metric does not describe the whole Poincare AdS, given that from Eq. (\ref{eq:132}) the coordinates $u_\pm$ are restricted to be positive. The horizon of the massless hyperbolic black hole $\rho_+=L$ can be written in Poincare coordinates as
\begin{equation}\label{eq:139}
 \left(z,u_\pm,\vec{y}\right)\big|_{\rm Horizon}=
 \left(\xi,0,\vec{y}\right)\ .
\end{equation}
This horizon intersects with the Poincare boundary $z=0$ exactly at the horizon of the Rindler space-time of the CFT, $u_\pm=0$. As $\xi$ increases we go further into the bulk without changing any of the other coordinates. For a two-dimensional boundary we can plot this in the diagram that is on the left of Fig. \ref{fig:4}. In this figure, the massless hyperbolic black hole only describes the bulk region to the right of the horizon. In App. \ref{app:entropy} we will use this to give a proof of the RT holographic formula for the entanglement entropy of the half-space.

\subsubsection*{Ball region}

A completely analogous construction can be made for the ball region by starting from the massless hyperbolic black hole written as
$$
ds^2=
  -\left(\frac{\rho^2-L^2}{R^2}\right)d\tau^2+
  \left(\frac{L^2}{\rho^2-L^2}\right)d\rho^2+
  \rho^2\left(
  du^2+\sinh^2(u)ds^2_{S^{d-2}}
  \right)\ ,
$$
and considering the following bulk change of coordinates
\begin{equation}\label{eq:138}
z=
  \frac{RL}{
  \rho\cosh(u)+\sqrt{\rho^2-L^2}\cosh(\tau/R)
  }\ ,
  \quad 
w_\pm=R
  \frac{\rho\sinh(u)\pm
  \sqrt{\rho^2-L^2}
  \sinh(\tau/R)}
  {\rho\cosh(u)+\sqrt{\rho^2-L^2}\cosh(\tau/R)}
  \ ,
\end{equation}
where $w_\pm=r\pm t$. As we take the boundary limit $\rho\rightarrow +\infty$ one can check that the conformal transformation in Eq. (\ref{eq:103}) is recovered. After some work, the inverse bulk transformation can be found and written as
$$\rho=
  \frac{L}{2Rz}
  \sqrt{(R+w_+)(R+w_-)+z^2}
  \sqrt{(R-w_+)(R-w_-)+z^2}\ ,$$
$$\tanh(\tau/R)=
  \frac{R(w_+-w_-)}{R^2-(w_+w_-+z^2)}\ ,
  \qquad \qquad
  \tanh(u)=\frac{R(w_++w_-)}
  {R^2+(w_+w_-+z^2)}$$
Applying this change of coordinates to the massless hyperbolic black hole we find the AdS Poincare metric
$$
ds^2=
  \left(\frac{L}{z}\right)^2
  \left(dz^2
  -dt^2+dr^2+r^2ds^2_{S^{d-2}}
  \right)\ .
$$
Same as before, the Poincare section is not covered entirely. To see what amount is covered by the black hole we can write the horizon $\rho_+=L$ in Poincare coordinates, which results in
\begin{equation}\label{eq:136}
\left(z,w_\pm,\theta_i\right)\big|_{\rm Horizon}=
  \left(\frac{R}{\cosh(u)},
  R\tanh(u),
  \theta_i
  \right)
  \qquad \Longrightarrow \qquad
  z^2+w_\pm^2=R^2\ ,
\end{equation}
where $\theta_i$ are the angles parametrizing $S^{d-2}$. As $u\rightarrow +\infty$ we approach the boundary of the Poincare AdS, where the black hole horizon goes to $w_\pm \rightarrow R$ that is precisely the boundary of the causal domain of the ball. For a two-dimensional CFT we can plot the surface in the diagram that is on the right of Fig. \ref{fig:4}.

This construction was famously developed in Ref. \cite{Casini:2011kv} to compute the entanglement entropy of the Minkowski ground state reduced to a ball and provide an explicit proof of the RT holographic prescription. In App. \ref{app:entropy} we show how it can be very naturally extended to give proof of the RT formula for the half-space region.

\section{Final remarks}
\label{sec:Dis}

In this work we have constructed localized thermal states with negative energy density in the causal domain of a ball and the half-space of a CFT in Minkowski space-time. We obtained explicit expressions for the energy density at arbitrary temperatures for a variety of CFTs and space-time dimensions. As the temperature changes these states interpolate between positive and negative energy without violating any of the QEIs present in the literature. In fact, for any two-dimensional CFT we have shown that the zero temperature solution saturates the most general QEI.

It would be interesting to extend these results by constructing localized states in other CFTs. This can be done for a CFT in a cylindrical background $\mathbb{R}\times S^{d-1}$ by taking the region given by a cap $\theta\in[0,\theta_0]$ at $t=0$, which allows for a local modular flow. From this we can define a localized thermal state which can be mapped to a thermal state in $\mathbb{R}\times \mathbb{H}^{d-1}$ \cite{Casini:2011kv}, so that its energy density can be obtained in the same way as for the ball and half-space in Minkowski. Another setup which allows for the same construction is the static de Sitter patch of a CFT in de Sitter space-time \cite{Candelas:1978gf,Casini:2011kv}.

An important feature of all these localized states which makes them appealing is that they can  be mapped via a conformal transformation and the AdS/CFT correspondence to black holes with hyperbolic horizon. This provides a tractable setup in which negative energy states can be studied in strongly coupled CFTs, which is very challenging using standard field theory tools. The well known negative mass of the hyperbolic black hole (which is very peculiar from the gravitational perspective) is nothing more than the manifestation of the negative energy density of the boundary field theory. 

In the process of investigating the holographic description of the localized thermal states we have found an explicit proof of the Ryu-Takaganayi holographic entropy formula for the vacuum reduced to the half-space. This gives a natural extension of the CHM proof for ball regions presented in Ref. \cite{Casini:2011kv}. Given the wide range of applications found to the CHM map, one might wonder whether this generalized construction might be of value in similar setups. 

For instance, in Ref. \cite{Johnson:2018amj} the CHM mapping was used in order to establish a concrete connection between extended black hole thermodynamics (see Ref. \cite{Kubiznak:2016qmn} for a review) and renormalization group flows in the boundary field theory. It would be interesting to see how this picture can be enlarged when considering the mapping to the half-space instead. Moreover, it might also be valuable to apply this holographic construction to study the entanglement entropy of the half-space under geometric and state deformations. This would be particurlarly interesting given that universal terms are known to arise when considering these types of deformations \cite{Rosenhaus:2014zza,Faulkner:2015csl}.

Using the localized state constructed in this paper it is a simple matter to consider separable states that are defined in the whole Hilbert space. For instance, if we split the $t=0$ Cauchy surface in Minkowski in the left and right half-space, the Hilbert space becomes $\mathcal{H}=\mathcal{H}_L\otimes \mathcal{H}_R$ and we can consider the following global state
\begin{equation}\label{eq:164}
\rho_{\rm sep}(\beta_L,\beta_R)=
  \rho_{HS}(\beta_L)\otimes \rho_{HS}(\beta_R)\ ,
\end{equation}
where $\rho_{HS}(\beta)$ is given in Eq. (\ref{eq:112}). This might seem worrying given that if we tune the left and right inverse temperatures so that $\beta_{L/R}>2\pi \alpha$ the energy density is negative in both sides. In fact, the total energy appears to diverges to minus infinity $E(\rho_{\rm sep})\rightarrow -\infty$, in direct contradiction with countless of results in the literature. 

This is resolved by analyzing the energy density of $\rho_{\rm sep}$ precisely at the entangling surface. In App. \ref{app:Juan} we do so for a separable state constructed from the reduced density operators in the left and right side and show that a positive and divergent contribution arises at the surface due to short distance correlations. This shows that the construction of a separable state such as the one in Eq. (\ref{eq:164}) requires and infinite amount of positive energy at the entangling surface which must exceed the negative energy contributions present for $\beta_{L/R}>2\pi \alpha$. Further investigations of these separable states are left to future work.

\section*{Acknowledgements}

We are grateful for useful comments and/or discussions with Clifford V. Johnson, Juan Hernandez, Robert Walker, Horacio Casini, David Blanco, Robert C. Myers, Andrew Svesko and Cynthia A. Keeler. This work was partially supported by DOE grant DE-SC0011687.

\appendix
\addtocontents{toc}{\protect\setcounter{tocdepth}{1}}

\section{Rindler entanglement from bulk horizon area}
\label{app:entropy}

The entanglement entropy of the Minkowski vacuum reduced to the half-space is given by
\begin{equation}\label{eq:148}
S_{\rm EE}=-{\rm Tr}
  \big(\rho_{HS}\ln(\rho_{HS})\big)\ ,
  \qquad \qquad \rho_{HS}=
  \frac{1}{Z}
  \exp\left(-2\pi\alpha K_\eta\right)\ ,
\end{equation}
where $K_\eta$ is proportional to the boost generator in Eq. (\ref{eq:112}) and from the Unruh effect \cite{Unruh:1976db,Bisognano:1976za} $\rho_{HS}$ is the Minkowski vacuum reduced to Rindler. Since the Von Neumann entropy is invariant under unitary transformations, the entanglement entropy can be computed at any point of the mappings in Fig. \ref{fig:6}. 

In particular it can be obtained from the black hole horizon entropy of the massless hyperbolic black hole in Eq. (\ref{eq:143}). For Einstein gravity the horizon entropy is given by the Bekenstein-Hawking area law and can be explicitly written as
\begin{equation}\label{eq:137}
S_{\rm EE}=
  S_{\rm Horizon}=\frac{2\pi}{\ell_p^{d-2}}A_{\rm Horizon}=
  \frac{4\pi a_d^*}{{\rm Vol}(S^{d-1})}
  {\rm Vol}(\mathbb{H}^{d-1})\ ,
\end{equation}
where $\ell_p$ is Planck's length and we have used the expression for the generalized central charge $a_d^*$ in Einstein gravity (\ref{eq:79}). If instead we consider an arbitrary covariant theory of gravity, the hyperbolic massless black hole is unchanged (still given by Eq. (\ref{eq:143})) but the horizon entropy is computed from Wald's functional \cite{Wald:1993nt,Jacobson:1993vj,Iyer:1994ys}. Nonetheless, the end result is the same as in the Einstein case \cite{Casini:2011kv} in Eq. (\ref{eq:137}).

The volume of the unit hyperbolic plane ${\rm Vol}(\mathbb{H}^{d-1})$ is divergent and must be regularized. Depending on the regularization procedure we can recover either the entanglement entropy of the half-space or the ball. Despite the fact that the Von Neumann entropy is invariant under all the unitary transformations in Fig. \ref{fig:6}, the regularization procedure is not. Depending whether we are interested in the half-space or the ball, it is convenient to write the divergent volume from Eqs. (\ref{eq:161}) or (\ref{eq:163}) as
$${\rm Vol}(\mathbb{H}^{d-1})=
  \begin{cases} 
  \qquad \,\,\,\,
  \displaystyle
  \mathcal{A}_{d-2}\int_0^{+\infty}\frac{d\xi}{\xi^{d-1}}
  \qquad \quad\,\ ,
  & {\rm for\,\,half-space} \vspace{8pt}\\
  \displaystyle
  {\rm Vol}(S^{d-2})
  \int_0^{+\infty}
  du\sinh^{d-2}(u)\,\ ,
  &  {\rm for\,\,ball}\ , \\
 \end{cases}
$$
where $\mathcal{A}_{d-2}$ is the area of the Rindler entangling surface, an infinite $(d-2)$-dimensional plane. The remaining integrals must be regularized by adding the cut-offs $\xi_{\rm min}$ and $u_{\rm max}$. We can relate them to the cut-off in Poincare coordinates $z_{\rm min}$ by evaluating the bulk change of coordinates in  Eqs. (\ref{eq:132}) and (\ref{eq:138}) at the horizon $\rho_+=L$
$$\xi_{\rm min}=
 z_{\rm min}\ ,
  \qquad \qquad
  u_{\rm max}=
  {\rm arcosh}\left(
  R/z_{\rm min}\right)\ .$$
With this prescription we can solve the remaining integrals and regulate the hyperbolic volume. For the half-space we obtain the following expression
\begin{equation}\label{eq:145}
{\rm Vol}(\mathbb{H}^{d-1})\big|_{HS}=
  \begin{cases} 
  \,\,\,\,\,\,\,\,\,\ln(\Lambda/\epsilon)
  \,\,\,\,\,\ ,
  & {\rm for\,\,}d=2  \vspace{6pt}\\
  \,\,
  \dfrac{\mathcal{A}_{d-2}}
  {(d-2)\epsilon^{d-2}}\ ,
  &  {\rm for\,\,}d>2\ , \\
 \end{cases}
\end{equation}
where we have defined $\epsilon=z_{\rm min}$ and for the $d=2$ case we had to introduce and additional long distance cut-off $\Lambda$. An analogous expression can be easily written for the case of the ball \cite{Hung:2011nu}. Using this in Eq. (\ref{eq:137}) we obtain the entanglement entropy of the Minkowski vacuum reduced to Rindler
\begin{equation}\label{eq:144}
S_{\rm Horizon}=
  \begin{cases} 
  \qquad \,\,\,\,\,\,\,\,
  \dfrac{c}{6}
  \ln(\Lambda/\epsilon)
  \quad \,\,\,\,\,\,\,\,\,\,\, \ ,
  & {\rm for\,\,}d=2 \vspace{7pt}
  \\ 
  \,\,
  \dfrac{4\pi a_d^*}{(d-2){\rm Vol}(S^{d-1})}
  \dfrac{\mathcal{A}_{d-2}}{\epsilon^{d-2}}
  \ ,
  & {\rm for\,\,}d>2\ , \\
 \end{cases}
\end{equation}
where we have used that $a_2^*=c/12$ with $c$ the Virasoro central charge \cite{Brown:1986nw,Myers:2010xs}. Due to the simple geometry of the entangling surface, the entanglement entropy contains no universal terms apart from the two-dimensional case. This means that the coefficient of the area term is regularization dependent. However, for a fixed regularization procedure the entanglement entropy computed via any other procedure will be equivalent. We will use this fact when comparing with the RT formula in the following section.

This result was obtained from the horizon area of the massless hyperbolic black hole at fixed time, that in Poincare coordinates is given by the bulk surface in Eq. (\ref{eq:139})
$$\left(z,t,x,\vec{y}\right)
  \big|_{\rm Horizon}=\left(
  \xi,0,0,\vec{y}
  \right)\ ,$$
with $\xi\ge 0$. In the $d=2$ case it corresponds to the highlighted line in the diagram that is on the left of Fig. \ref{fig:4}, that goes straight into the bulk. For larger dimensions the picture cannot be drawn but is essentially unchanged given that the transverse coordinates $\vec{y}$ play no fundamental role.

\subsection{Matching with RT prescription}
\label{app:RT}

We now compute the entanglement entropy of the Minkowski vacuum reduced to Rindler from the Ryu-Takayanagi prescription \cite{Ryu:2006bv,Ryu:2006ef}, that is given by
\begin{equation}\label{eq:142}
S_{\rm RT}=\frac{2\pi}{\ell_p^{d-1}}
  {\rm ext}\left[A(\gamma)\right]
  \ ,
\end{equation}
where ${\rm ext}\left[A(\gamma)\right]$ is the area of the extremal bulk surface $\gamma$ whose boundary lies on the entangling surface. While the calculation from the previous section holds for an arbitrary theory of gravity, this formula is only expected to be true for holographic CFTs dual to Einstein gravity.

Since the global state is the Minkowski vacuum, we must consider the bulk extremal surface in the Poincare metric (\ref{eq:140}) at $t=0$. From the symmetry of the setup, we take an ansatz for the surface that is independent of $\vec{y}$, so that its induced area becomes
\begin{equation}\label{eq:141}
A(\gamma)=
  L^{d-1}
  \int d\vec{y}\int 
  \frac{\sqrt{dz^2+dx^2}}{z^{d-1}}
  \ .
\end{equation}
The $\vec{y}$ integral gives the area of the entangling surface $\mathcal{A}_{d-2}$, while the remaining is the functional we must extremize.

\begin{figure}
\centering
\includegraphics[scale=0.53]{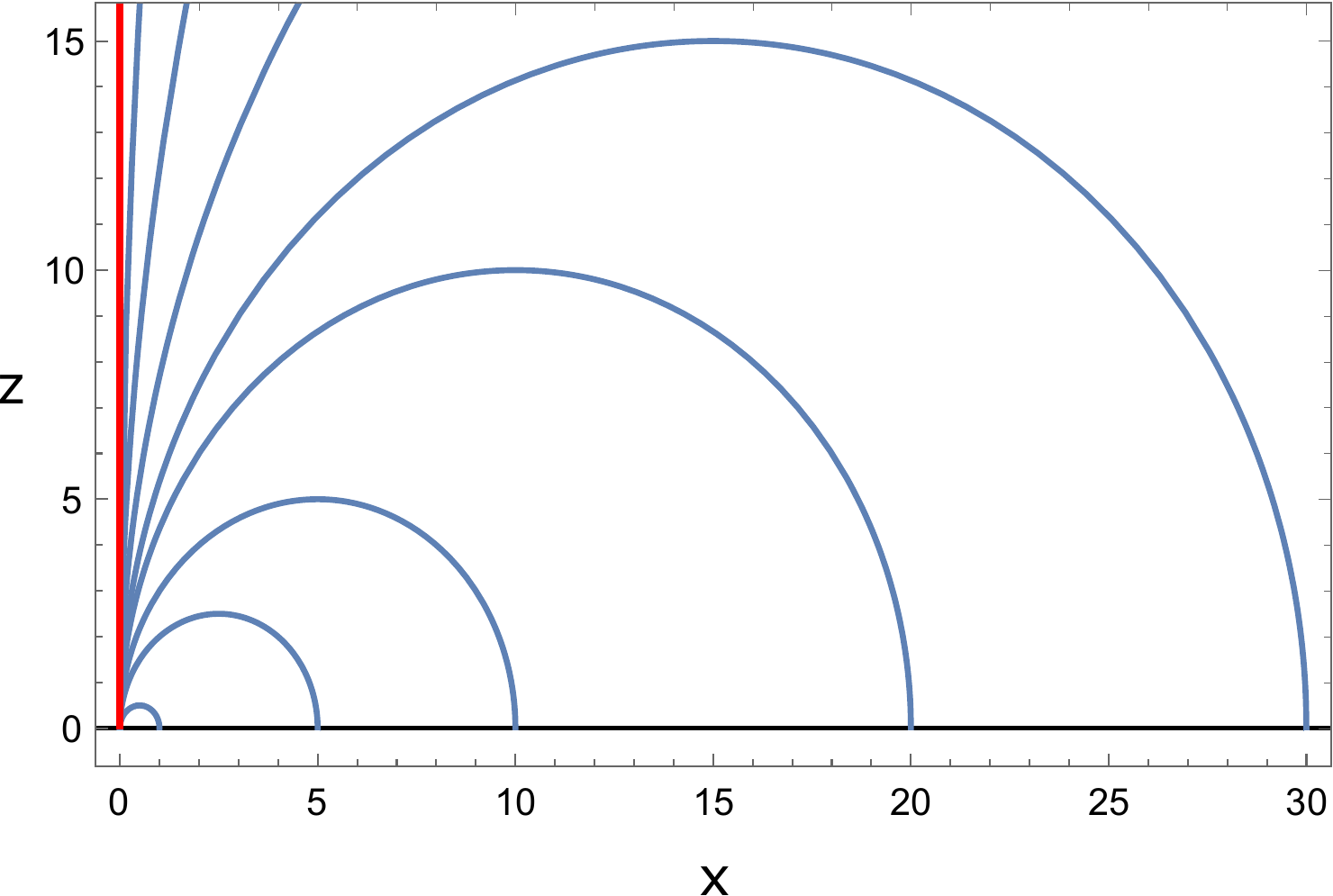}
\caption{Ryu-Takayanagi extremal curves for CFT$_2$/AdS$_3$ given by semi-circles $z^2+(x-\Lambda)^2=\Lambda^2$ with $z>0$. In the limit of $\Lambda\rightarrow +\infty$ we see how they limit towards the red vertical line, that is precisely the horizon at $t=0$ in the diagram that is on the left of Fig. \ref{fig:4}.}\label{fig:7}
\end{figure}

Let us first consider the $d=2$ case where the half-space is given by a semi-infinite segment in the boundary. If we add a long distance regulator $\Lambda$, the segment becomes finite and given by ${x\in [0,2\Lambda]}$. The curve extremizing the functional (\ref{eq:141}) can be easily computed \cite{Ryu:2006bv} and is given by the semi-circle ${z^2+(x-\Lambda)^2=\Lambda^2}$ with $z>0$. Evaluating the area functional (\ref{eq:141}) along this curve we find
\begin{equation}\label{eq:30}
{\rm ext}\left[A(\gamma)\right]=
  2L\int_\epsilon^{\Lambda}
  \frac{dz}{z}\sqrt{1+x'(z)}=
  2L\ln(\Lambda/\epsilon)\ ,
\end{equation}
where we have regularized the integral the same way as in Eq. (\ref{eq:145}), by adding the cut-off $z_{\rm min}=\epsilon$. The factor of 2 takes into account the area of both halves of the semi-circle. However, as we take the limit $\Lambda\rightarrow +\infty$ we effectively get half of a semi-circle, meaning that we must drop this extra factor (see Fig. \ref{fig:7}). Using this in the RT formula (\ref{eq:142}), we find
$$S_{\rm RT}=\frac{2\pi L}{\ell_p}
  \ln(\Lambda/\epsilon)=
  \frac{c}{6}
  \ln(\Lambda/\epsilon)\ ,$$
where in the last equality we have written the result in terms of the Virasoro central charge $c$. Comparing with the result obtained from the black hole horizon area in Eq. (\ref{eq:144}), we find precise agreement.

We can compare the RT extremal curve with the horizon of the massless hyperbolic black hole. In Fig. \ref{fig:7} we have plotted the extremal curve for increasing values of $\Lambda$, which limits towards the red vertical line at $x=0$ as $\Lambda\rightarrow +\infty$. Comparing with the previous section, this red vertical line becomes the bulk horizon at ${t=0}$, highlighted in the diagram that is on the left of Fig. \ref{fig:4}. This means that the Ryu-Takayanagi extremal curve precisely matches with the horizon of the massless hyperbolic black hole. Given that the coordinates $\vec{y}$ do not play any essential role, this generalizes to arbitrary dimensions where it is not possible to draw the picture.

To evaluate the area of the RT surface for general $d$ we use the result obtained for a strip ${x\in[-\Lambda,\Lambda]}$, that is given by \cite{Ryu:2006ef}
$${\rm ext}
  \left[A(\gamma_{\rm strip})\right]=
  \frac{2L^{d-1}}{d-2}
  \frac{\mathcal{A}_{d-2}}{\epsilon^{d-2}}+
  \mathcal{O}(1/\Lambda^{d-2})\ ,$$
where $\epsilon=z_{\rm min}$, the same regulator.\footnote{Ref. \cite{Giataganas:2019wkd} has recently used this result from the RT formula to compute the Rindler entanglement entropy.} Similarly to the two-dimensional case, we must take the limit $\Lambda\rightarrow +\infty$ and divide by a factor of two to account for the area excess. Doing so and using this in Eq. (\ref{eq:142}) we find
$$S_{\rm RT}=
  \frac{2\pi L^{d-1}}{(d-2)\ell_p^{d-1}}
  \frac{\mathcal{A}_{d-2}}{\epsilon^{d-2}}=
  \frac{4\pi a_d^*}{(d-2){\rm Vol}(S^{d-1})}
  \frac{\mathcal{A}_{d-2}}{\epsilon^{d-2}}
  \ ,$$
where in the last equality we have used the expression of $a_d^*$ for Einstein gravity in Eq. (\ref{eq:79}). Comparing with Eq. (\ref{eq:144}) we find precise agreement. This allows us to conclude that 
$$S_{\rm EE}=S_{\rm Horizon}=S_{\rm RT}\ ,$$
for the Minkowski vacuum reduced to the half-space of a CFT, providing with a proof of the RT prescription in this particular case.

\subsection{Renyi entropy}
\label{app:Renyi}

By a slight modification of this approach we can also compute the Renyi entropy, defined as
$$S_q=\frac{\ln\left[
  {\rm Tr}\left(\rho_{HS}^q\right)
  \right]}{1-q}\ ,$$
where in the limit $q\rightarrow 1$ we recover the entanglement entropy. This expression can be rewritten in terms of the free energy of the thermal state in $\mathbb{R}\times \mathbb{H}^{d-1}$, $\bar{F}(\beta)$ as \cite{Baez,Hung:2011nu}
\begin{equation}\label{eq:149}
S_q=\left(\frac{q}{1-q}\right)2\pi\alpha
  \left[
  \bar{F}(2\pi \alpha)-\bar{F}(2\pi \alpha q)
  \right]\ .
\end{equation}
An important difference with respect to the entanglement entropy is that the Renyi entropy involves the thermal state in the hyperbolic space-time at any temperature. For the holographic calculation this means that we must consider hyperbolic black holes solutions away from the massless case. Since the solution for $\beta\neq 2\pi\alpha$ depends on the gravity theory under consideration, the Renyi entropy will differ in each case. 

Precisely the calculation of Eq. (\ref{eq:149}) was considered in Ref. \cite{Hung:2011nu} for several gravity theories. For Einstein gravity the result is given by \cite{Hung:2011nu}
\begin{equation}\label{eq:150}
S_q=\left(\frac{q}{1-q}\right)
  \left[
  1-\frac{1}{2} x_q^{d-2}
  (1+x_q^2)
  \right]
  S_{\rm Horizon}\ ,
\end{equation}
where
$$x_q=\frac{1+\sqrt{1+q^2d(d-2)}}{qd}\ ,$$
and $S_{\rm Horizon}$ is given in Eq. (\ref{eq:137}). The only difference with respect to the results in Ref. \cite{Hung:2011nu} is the regularization of the hyperbolic volume, which in this case is given by Eq. (\ref{eq:145}). Keeping this in mind, one can borrow the results of Ref. \cite{Hung:2011nu} and obtain the Rindler Renyi entropy from more complicated gravity theories.

We also use Eq. (\ref{eq:149}) to compute the Renyi entropy directly in field theory. For two-dimensional CFTs the free energy can be obtained from the basic thermodynamic relation in Eq. (\ref{eq:47}) and the energy density obtained in Eq. (\ref{eq:105}), so that we find
$$\bar{F}(\beta)=
  -\frac{c\pi }{6\beta^2}\,\alpha
  {\rm Vol}(\mathbb{H})+\bar{F}_0
  $$
Using this in Eq. (\ref{eq:149}) and the regulated hyperbolic volume in Eq. (\ref{eq:145}) we find the Rindler Renyi entropy for an arbitrary two-dimensional CFT
$$S_q=
  \frac{c}{12}
  \left(1+\frac{1}{q}\right)
  \ln(\Lambda/\epsilon)\ ,$$
that agrees with the holographic results in Eqs. (\ref{eq:144}) and (\ref{eq:150}). Moreover, it agrees with the calculations in Refs. \cite{Callan:1994py,Kabat:1994vj,Dowker:1994fi} for the entanglement entropy of a free scalar and fermion. 

For a massless scalar field we can obtain $S_q$ from the free energy computed in App. \ref{app:hyp}. For even space-time dimensions we can use the expressions of Table \ref{table:4} in the Eq. (\ref{eq:149}) to obtain the Renyi entropies in Table \ref{table:2}.
\begin{table*}\setlength{\tabcolsep}{3pt}
\centering
\begin{tabular}{|Sc|Sc|Sc|Sc|Sc|}
\hline 
$d$ & 4 & 6 & 8  \\  \hline
$S_q$  &
  $\displaystyle
  \frac{(1+q)(1+q^2)}{1440\pi q^3}
  \frac{\mathcal{A}_2}{\epsilon^2}
  $  &
  $\displaystyle
  \frac{(1+q)(1+3q^2)(2+3q^2)}
  {120960\pi^2q^5}
  \frac{\mathcal{A}_4}{\epsilon^4}$  &
  $\displaystyle
  \frac{(1+q)(3+23q^2+79q^4+79q^6)}{3628800\pi^3 q^7}
  \frac{\mathcal{A}_6}{\epsilon^6}
  $   \\ \hline
\end{tabular}
\caption{Renyi entropy of the Minkowski vacuum reduced to the half-space for a free massless scalar.}\label{table:2}
\end{table*}
Taking the $q\rightarrow 1$ limit for $d=4$ we recover the entanglement entropy computed in Ref. \cite{Susskind:1994sm}. We also find agreement with the holographic result (\ref{eq:144}), using that from Eq. (\ref{eq:120}) $a_4^*$ is given by the trace anomaly of a real scalar ${A_4=1/360}$. If we compare the Renyi entropies for arbitrary $q$ in Table \ref{table:2} with the holographic result in Eq. (\ref{eq:150}), we find that the expressions do not agree. This is not a problem since the holographic calculations are expected to hold for strongly coupled CFTs with a large number of degrees of freedom, which is clearly not the case of a free scalar field.

For $d=3$ space-time dimensions we can use the integral expression for the free energy in Eq. (\ref{eq:71}) and write the Renyi entropy as
$$
\begin{aligned}
S_q=
  \left(\frac{q}{1-q}\right)
  \frac{1}
  {4\pi\sqrt{2}}
  \frac{\mathcal{A}_1}{\epsilon}
  \int_0^\infty		&	
  \frac{dv\,\sinh^2(v/2)}{(v/2)^2\left(\cosh(v)-1\right)^{5/2}}
  \,\,\times
  \\
   &\left[
  \left(
  \frac{\sinh(v/2)}{q\sinh(v/2q)}
  \right)^2
  \left(v+q\sinh(v/q)\right)
  -
  \left(v+\sinh(v)\right)
  \right]\ .
\end{aligned}
$$
Using numerical integration we can solve the integral and obtain the Renyi entropy as a function of $q$. In the limit of $q\rightarrow 1$ we obtain the entanglement entropy
\begin{equation}\label{eq:151}
S_{q\rightarrow 1}=S_{EE}\approx 0.0102\frac{\mathcal{A}_1}{\epsilon}\ .
\end{equation}
Comparing with the holographic result in Eq. (\ref{eq:144}) we see that the area terms is exactly given by $a_3^*$. The value of $a_3^*$ is computed from  Eq.(\ref{eq:120}) using the expression for $\ln[Z(S^3)]$ obtained in Eq. (2.5) of Ref. \cite{Pufu:2016zxm}, which gives $a_3^*\approx 0.0102$ in precise agreement with Eq. (\ref{eq:151}).

The fact that all the different calculations of the entanglement entropy agree serves a check of our results. Despite of the fact that the coefficient of the area term is not universal all our results agree and are consistent because we are always using the same regularization procedure for the hyperbolic volume. 

There are several other calculations one could perform regarding the Rindler Renyi entropy. For instance, it would be interesting to compare this approach with the holographic formula for the Renyi entropy proposed in Ref. \cite{Dong:2016fnf} Moreover, one could consider the charged \cite{Belin:2013uta} and extended \cite{Johnson:2018bma} Renyi entropy for the half-space and analyze the analiticity as a function of $q$ by perfoming a similar analysis to the one in Ref. \cite{Belin:2013dva}.

\section{Conformal scalar in hyperbolic space-time}
\label{app:hyp}

In this Appendix, we apply standard canonical quantization on a scalar field conformally coupled to the hyperbolic space-time and compute the thermal energy density. The systematic canonical quantization of a thermal scalar field in $\mathbb{R}\times \mathbb{H}^{d-1}$ has not been presented in the literature, though substancial and important work has been previously done in Refs. \cite{Bunch:1978ka,Candelas:1978gf,Denardo:1981xa,Brown:1982hb,Pfautsch:1982hv,Page:1982fm,Bytsenko:1994bc,Moretti:1995fa,Iellici:1997yh,Haba:2007ay,Cho:2014ira,Klebanov:2011uf}. We explicitly do so for arbitrary temperature and space-time dimensions, and compute the thermal two-point function, energy density and partition function.

The action of the scalar field is given by
\begin{equation}\label{eq:27}
S_{\rm scalar}=-\frac{1}{2}
  \int d^dX\,
  \sqrt{-\bar{g}}
  \Big(
  \bar{g}^{ab}
  (\partial_a \bar{\phi})
  (\partial_b \bar{\phi})+
  \frac{(d-2)}{4(d-1)} \bar{\mathcal{R}}\bar{\phi}^2
  \Big)\ ,
\end{equation}
where $X^a=(\eta,u,\theta_i)$ and we find convenient to write the space-time metric $\mathbb{R}\times \mathbb{H}^{d-1}$ as
\begin{equation}\label{eq:162}
d\bar{s}^2=-d\eta^2+\alpha^2\left(
  du^2+\sinh^2(u)ds^2_{S^{d-2}}
  \right)\ .
\end{equation}
From this we can easily compute the Ricci tensor and scalar, that are given by
\begin{equation}\label{eq:51}
\bar{\mathcal{R}}^a_{\,\,\,b}=
  -\frac{(d-2)}{\alpha^2}
  {\rm diag}\left(
  0,1,\dots,1
  \right)\ ,
  \qquad \qquad
  \bar{\mathcal{R}}=-\frac{(d-2)(d-1)}{\alpha^2}\ .
\end{equation}
We have chosen the normalization of the action such that the canonical momentum $\pi(X)$, is properly normalized and given by $\bar{\pi}(X)=\partial_\eta\bar{\phi}(X)$. Upon variation of the action, we obtain the following wave equation
\begin{equation}\label{eq:12}
\left[
  \bar{\nabla}^2+
  \left(
  \frac{d-2}{2\alpha}
  \right)^2
  \right]\bar{\phi}(X)=0\ ,
\end{equation}
where $\bar{\nabla}^2$ is the Laplacian operator of the metric in Eq. (\ref{eq:162}). The differential equation is separable and can be solved explicitly. In App. \ref{app:hypeq} we show that it can be solved by the following mode solution
\begin{equation}\label{eq:31}
H_{\vec{k}}(X)=
 e^{-iw\eta/\alpha}Y^m_\ell(\theta_i)f_{w,\ell}(u)
 \ ,
\end{equation}
where $\vec{k}=(w,\ell,m)$, $Y^m_\ell(\theta_i)$ are the spherical harmonics on $S^{d-2}$ and the functions $f_{w,\ell}(u)$ are explicitly written in App. \ref{app:hypeq}. The coefficient $w$ is real and non-negative, while $\ell$ and $m$ are integers obeying the standard constraints. Since these modes form a complete and orthogonal set (see App. \ref{app:hypeq} for details) we can use them to expand $\bar{\phi}(X)$ and $\bar{\pi}(X)$. Applying standard canonical quantization, we promote the classical fields to operators in the Hilbert space and find
\begin{equation}\label{eq:21}
\bar{\phi}(X)=\alpha^{\frac{2-d}{2}}
  \sum_{\ell,m}
  \int_0^{\infty}\frac{dw}{\sqrt{2w}}
  \left[
  \bar{a}_{\vec{k}}\,
  H_{\vec{k}}(X)+
  \bar{a}^\dagger_{\vec{k}}\,
  H^*_{\vec{k}}(X)
  \right]\ ,
\end{equation}
$$\bar{\pi}(X)=
  -i\alpha^{-d/2}
  \sum_{\ell,m}
  \int_0^\infty dw\sqrt{\frac{w}{2}}
  \left[
  \bar{a}_{\vec{k}}\,
  H_{\vec{k}}(X)-
  \bar{a}^\dagger_{\vec{k}}\,
  H^*_{\vec{k}}(X)
  \right]\ ,
  $$
where the normalization has been chosen so that the operators $\bar{a}_{\vec{k}}^\dagger$ and $\bar{a}_{\vec{k}}$ satisfy
$$\Big[\bar{a}_{\vec{k}},\bar{a}_{\vec{k}'}\Big]=
  \Big[\bar{a}_{\vec{k}}^\dagger,\bar{a}_{\vec{k}'}^\dagger\Big]=0 \ ,
  \qquad \qquad
  \Big[\bar{a}_{\vec{k}},
  \bar{a}_{\vec{k}'}^\dagger\Big]=
  \delta_{mm'}\delta_{\ell \ell'}
  \delta(w-w')\ ,$$
and $\bar{\phi}(X)$ and $\bar{\pi}(X)$ verify the standard equal time canonical commutation relation. This allows us to define the ground state (that may not be unique) associated to the hyperbolic space-time as $\bar{a}_{\vec{k}}\ket{0_{\rm Hyp}}=0$. The rest of the Hilbert space is generated by acting with an arbitrary number of creation operators on $\ket{0_{\rm Hyp}}$.

\subsection{Thermal two point function}

To compute the thermal free energy we must first obtain the thermal two point function of the operator $\bar{\phi}(X)$, that is given by
\begin{equation}\label{eq:41}
\bar{G}_\beta(X,X')= \frac{1}{Z}
  {\rm Tr}\left(
  e^{-\beta \bar{H}_\eta}
  \bar{\phi}(X)
  \bar{\phi}(X')
  \right)\ ,
\end{equation}
where $Z$ is the thermal partition function and $\bar{H}_\eta$ is the Hamiltonian in the hyperbolic space-time generating $\eta$ translations. Instead of computing the trace in Eq. (\ref{eq:41}) directly, we note that from the KMS condition \cite{Haag:1992hx} the thermal two point function is periodic with period $\beta$ in the imaginary time direction \textit{i.e.}, $\bar{G}_\beta(X,X')=\bar{G}_\beta(X+i\beta X^0,X')$. We can then start from the vacuum two point function $\bar{G}_0(X,X')$, and use the method of images to impose the required periodicity \cite{Birrell:1982ix}
\begin{equation}\label{eq:14}
\bar{G}_{\beta}(X,X')=
  \sum_{n\in \mathbb{Z}}
  \bar{G}_0(X,X')\Big|_{\eta\rightarrow \eta+in\beta}\ .
\end{equation}
This gives a simple procedure for computing the thermal two point function when considering free theories. The vacuum two point function can be easily obtained from the expansion of $\bar{\phi}(X)$ in Eq. (\ref{eq:21}) and the algebra of the creation/annihilation operators, so that we find
\begin{equation}\label{eq:13}
\bar{G}_{0}(X,X')=
  \bra{0_{\rm Hyp}}
  \bar{\phi}(X)
  \bar{\phi}(X')
  \ket{0_{\rm Hyp}}=
  \alpha^{2-d}
  \int_0^{\infty}\frac{dw}{2w}
  \sum_{\ell,m}
  H_{\vec{k}}(X)
  H^*_{\vec{k}}(X')\ .
\end{equation}
The summation over $\ell$ and $m$ can be explicitly solved, but the result is very different depending on whether $d$ is even or odd. Because of this, we must consider each case separately.

\subsubsection*{Even space-time}

The vacuum two point function (\ref{eq:13}) can be written explicitly with the aid of the following identity \cite{Bander:1965im}
$$\sum_{\ell,m}
  H_{\vec{k}}(X)
  H^*_{\vec{k}}(X')=
  \frac{2we^{-iw\Delta \eta/\alpha}}
  {(-2\pi)^{d/2}}
  \left(
  \frac{d}{d\cosh(\Delta \gamma)}
  \right)^{\frac{d-4}{2}}
  \frac{\sin(w\Delta \gamma)}{\sinh(\Delta \gamma)}
  \ ,$$
where $\Delta \eta=\eta-\eta'$ and $\Delta \gamma$ is defined from
\begin{equation}\label{eq:16}
\cosh(\Delta \gamma)=\cosh(u)\cosh(u')-
  \sinh(u)\sinh(u')\hat{n}.\hat{n}'\ ,
\end{equation}
with $\hat{n}$ the unit vector of $S^{d-2}$ given from $X^a$. Though $\Delta \gamma$ has in general a complicated expression in terms of the spatial coordinates $(u,\theta_i)$, we can always align our coordinate system on $S^{d-2}$ so that $\hat{n}.\hat{n}'=1$ which implies $\Delta \gamma=\Delta u$. For the moment we will keep $\Delta \gamma$ in full generality. Using this identity, the vacuum two point function in Eq. (\ref{eq:13}) becomes
$$\bar{G}_{0}(X,X')=
  \frac{\alpha^{2-d}}{(-2\pi)^{d/2}}
  \left(
  \frac{d}{d\cosh(\Delta \gamma)}
  \right)^{\frac{d-4}{2}}
  \frac{1}{\sinh(\Delta \gamma)}
  \int_0^{\infty}dw
  e^{-iw\Delta \eta/\alpha}
  \sin(w\Delta \gamma)\ .$$
The resulting $w$ integral is divergent since it is given by an oscillatory function integrated over the real positive line. We can regulate such divergence by adding a small imaginary time component $\Delta\eta\rightarrow \Delta \eta- i\epsilon$ so that we get a damping exponential and find
\begin{equation}\label{eq:42}
\bar{G}_{0}(X,X')=
  \frac{\alpha^{4-d}}{(-2\pi)^{d/2}}
  \left(
  \frac{d}{d\cosh(\Delta \gamma)}
  \right)^{\frac{d-4}{2}}
  \frac{\Delta \gamma}{\sinh(\Delta \gamma)}
  \frac{1}
  {-\Delta \eta^2+\alpha^2\Delta \gamma^2+i\epsilon}\ .
\end{equation}
This expression should be understood as a distribution, which is no surprise since $\bar{\phi}(X)$ is an operator valued distribution. 

The thermal two point function can be obtained from the method of images by using Eq. (\ref{eq:14}). To do so, we first drop the factor $i\epsilon$ and use that the series can be summed exactly according to
\begin{equation}\label{eq:18}
\sum_{n\in \mathbb{Z}}
  \frac{1}
  {-\left(\Delta \eta+in\beta\right)^2+\alpha^2\Delta \gamma^2}=
  \frac{T}{2\alpha^2\Delta \gamma}
  \frac{\sinh(T\Delta \gamma)}
  {\cosh(T\Delta\gamma)-
  \cosh(T \Delta\eta/\alpha)}\ ,
\end{equation}
where on the right hand side we have defined the dimensionless temperature $T=2\pi \alpha/\beta$. Using this and Eq. (\ref{eq:42}) in Eq. (\ref{eq:14}), we obtain the following expression for the thermal two point function
\begin{equation}\label{eq:24}
\bar{G}_{\beta}(X,X')=
  \frac{\alpha^{2-d}}{2(-2\pi)^{d/2}}
  \left(
  \frac{d}{d\cosh(\Delta \gamma)}
  \right)^{\frac{d-4}{2}}
  \frac{\sinh(T\Delta \gamma)}{\sinh(\Delta \gamma)}
  \frac{T}
  {\cosh(T\Delta\gamma)-
  \cosh(T \Delta\eta/\alpha+i\epsilon)}\ ,
\end{equation}
where we have introduced the $i\epsilon$ factor again. For any particular space-time dimension the derivative can be easily computed and we can write the thermal two point function explicitly.\footnote{The way in which we have dealt with the $i\epsilon$ prescription and the method of images is sloppy but gives the right answer. The proper way is the following: first Wick rotate the vacuum two point function in Eq. (\ref{eq:42}) to Euclidean signature $\Delta\eta\rightarrow i\Delta\eta_e$. The Euclidean correlator does not require the $i\epsilon$ for convergence, so we can set it to zero as long as the operators are time ordered, \textit{i.e.} $\Delta\eta_e=\eta_e-\eta_e'>0$. We can then solve the summation over the images and finally analytically continue to back Lorentzian time $\Delta \eta_e=\epsilon-i\Delta \eta$, which gives the result in Eq. (\ref{eq:24}). We thank David Simmons-Duffin for clarifying this issue.}  

For the particular case in which $T=1$ (that corresponds to $\beta=2\pi \alpha$) the expression only depends on $\Delta \gamma$ through $\cosh(\Delta \gamma)$, which means we can solve the derivative for arbitrary values of $d$ and find
\begin{equation}\label{eq:19}
\bar{G}_{\beta=2\pi \alpha}(X,X')=
  \frac{\Gamma\big((d-2)/2\big)}
  {4\pi^{d/2}}
  \left[
  \frac{1}{2\alpha^2
  \left(\cosh(\Delta \gamma)
  -\cosh(\Delta\eta/\alpha+i\epsilon)
  \right)}
  \right]^{\frac{d-2}{2}}
  \ .
\end{equation}

\subsubsection*{Odd space-time}

For odd space-time dimension, we can write an expression for the vacuum two point function (\ref{eq:13}) by using the following identity \cite{Bander:1965im}
\begin{equation}\label{eq:17}
\sum_{\ell,m}
  H_{\vec{k}}(X)
  H^*_{\vec{k}}(X')=-
  e^{-iw\frac{\Delta \eta}{\alpha}}
  \frac{w\tanh(\pi w)}{(-2\pi)^{\frac{d-1}{2}}}
  \left(
  \frac{d}{d\cosh(\Delta \gamma)}
  \right)^{\frac{d-3}{2}}
  P_{iw-1/2}(
  \cosh(\Delta \gamma))
  \ ,
\end{equation}
where $\Delta \gamma$ is again defined from Eq. (\ref{eq:16}), and $P_{iw-1/2}(z)$ is the associated Legendre function \cite{Gradshteyn,Trans}. Using the following integral representation for $P_{iw-1/2}(z)$ \cite{Bander:1965im}
$$P_{iw-1/2}(
  \cosh(\Delta \gamma)
  )=
  \frac{\sqrt{2}}{\pi \tanh(\pi w)}
  \int_{\Delta \gamma}^\infty
  \frac{\sin(wv)dv}{\sqrt{\cosh(v)-\cosh(\Delta \gamma)}}\ ,$$
we can write the vacuum two point function in Eq. (\ref{eq:13}) as
$$\bar{G}_{0}(X,X')=
  \frac{\sqrt{2}\alpha^{2-d}}
  {(-2\pi)^{\frac{d+1}{2}}}
  \left(
  \frac{d}{d\cosh(\Delta \gamma)}
  \right)^{\frac{d-3}{2}}  
  \int_{\Delta \gamma}^\infty
  \frac{dv}{\sqrt{\cosh(v)-\cosh(\Delta \gamma)}}
  \int_0^{\infty}dw
  e^{-iw\Delta \eta/\alpha}
  \sin(wv)\ .$$
The $w$ integral is the same one we solved for the even $d$ case, so that we find
$$\bar{G}_{0}(X,X')=
  \frac{\sqrt{2}\alpha^{4-d}}
  {(-2\pi)^{\frac{d+1}{2}}}
  \left(
  \frac{d}{d\cosh(\Delta \gamma)}
  \right)^{\frac{d-3}{2}}  
  \int_{\Delta \gamma}^\infty
  \frac{dv\,v}{\sqrt{\cosh(v)-\cosh(\Delta \gamma)}}
  \frac{1}{-\Delta\eta^2+\alpha^2v^2+i\epsilon}\ .$$
Even though the remaining integral cannot be solved analytically, we can use Eqs. (\ref{eq:14}) and (\ref{eq:18}), to obtain the thermal two point function in terms of $T=2\pi \alpha/\beta$
\begin{equation}\label{eq:25}
\begin{aligned}
\bar{G}_{\beta}(X,X')=
  \frac{\alpha^{2-d}}
  {\sqrt{2}(-2\pi)^{\frac{d+1}{2}}}
  &\left(
  \frac{d}{d\cosh(\Delta \gamma)}
  \right)^{\frac{d-3}{2}}\\  
  &\int_{\Delta \gamma}^\infty
  \frac{dv}{\sqrt{\cosh(v)-\cosh(\Delta \gamma)}}
  \frac{T\sinh(Tv)}
  {\cosh(Tv)-
  \cosh(T \Delta\eta/\alpha+i\epsilon)}\ .
\end{aligned}
\end{equation}
We find that this expression can only be integrated analytically for several rational values of $T$. For the case in which $T=1$ (corresponding to $\beta=2\pi \alpha$) this can be easily done and we find
\begin{equation}\label{eq:43}
\bar{G}_{\beta=2\pi \alpha}(X,X')=
  \frac{\pi \alpha^{2-d}}
  {\sqrt{2}(-2\pi)^{\frac{d+1}{2}}}
  \left(
  \frac{d}{d\cosh(\Delta \gamma)}
  \right)^{\frac{d-3}{2}}  
  \frac{1}{\sqrt{\cosh(\Delta \gamma)-\cosh(\Delta\eta/\alpha+i\epsilon)}}\ .
\end{equation}
The remaining derivative can be computed explicitly for arbitrary values of $d$, and we find exactly the same result as in the even case in Eq. (\ref{eq:19}). For other integer values of $T$, the integral in Eq. (\ref{eq:25}) can also be solved exactly, though it becomes increasingly complicated. 

\subsubsection*{Thermal behavior of Minkowski vacuum}

For $\beta=2\pi \alpha$ we have found that the two point function is given by Eq. (\ref{eq:19}) for both even and odd $d$. This is not a coincide as we can see by applying the conformal transformations relating $\mathbb{R}\times \mathbb{H}^{d-1}$ to the causal domain of the ball in Minkowski, where the conformal factor is given in Eq. (\ref{eq:152}), \textit{i.e.} $\Omega(X)=\cosh(\eta/\alpha)+\cosh(u)$. Since the scaling dimension of a free scalar is $\Delta_\phi=(d-2)/2$, the two point function in the causal domain of the ball $G_{\beta}(X,X')$ transforms according to
$$G_{\beta}(X,X')=
  \bar{G}_{\beta}(X,X')\Omega^{-\Delta_\phi}(X)
  \Omega^{-\Delta_\phi}(X')\ .
  $$
Setting the coordinate system so that ${\Delta\gamma=\Delta u}$ and momentarily dropping the $i\epsilon$ prescription, we can use Eq. (\ref{eq:19}) to write the two point function at temperature $\beta=2\pi\alpha$ as
$$G_{\beta=2\pi \alpha}(X,X')=
  \frac{\Gamma\big((d-2)/2\big)}
  {4\pi^{d/2}}
  \left[
  \frac{\left(\cosh(u)+\cosh(\eta/\alpha)\right)
  \left(\cosh(u')+\cosh(\eta'/\alpha)\right)}{2\alpha^2
  \left(
  -\cosh(\Delta\eta/\alpha)+\cosh(\Delta u)
  \right)}
  \right]^{\frac{d-2}{2}}
  \ .$$ 
Using the change of coordinates in Eq. (\ref{eq:103}) (replacing $\tau\rightarrow \eta$ and $R \rightarrow \alpha$) to write this in Minkowski coordinates $(t,r,\theta_i)$ we find
$$G_{\beta=2\pi \alpha}(\Delta t,\Delta r)=
  \frac{\Gamma\big((d-2)/2\big)}
  {4\pi^{d/2}}
  \left[
  \frac{1}{-\Delta t^2+\Delta r^2+i\epsilon}
  \right]^{\frac{d-2}{2}}
  \ ,$$
that is precisely the Minkowski vacuum two point function \cite{Birrell:1982ix}. Moreover, since the theory is free and $\phi(X)$ is the only primary field, all the correlators are determined from this two point function using Wick's theorem. This means that for an observer restricted to $\mathcal{D}_\mathcal{B}$, the Minkowski vacuum seems to be a thermal state of inverse temperature $\beta=2\pi \alpha$. This gives a particular and explicit proof of the much more general result obtained in Ref. \cite{Casini:2011kv}.

\subsection{Thermal stress tensor and free energy}
\label{sec:stress}

We can now use the results from the previous section to compute the thermal expectation value of the stress tensor
\begin{equation}\label{eq:44}
\langle \bar{T}^a_{\,\,\,b} \rangle_\beta=
  \frac{1}{Z}
  {\rm Tr}
  \left(  
  e^{-\beta \bar{H}_\eta}
  \bar{T}^a_{\,\,\,b}
  \right)\ .
\end{equation}
The operator $\bar{T}^a_{\,\,\,b}$ is determined from the variation of the action in Eq. (\ref{eq:27}) with respect to the metric and can be written in terms of $\bar{\phi}$ and its derivatives as \cite{Parker:2009uva}\footnote{We must consider a different sign on the third term of Ref. \cite{Parker:2009uva} due to the signature convention. For the correct sign in four dimensions, see for example Ref. \cite{Meng:2016gyt}.}
\begin{equation}\label{eq:23}
\bar{T}^a_{\,\,\,b}=
  \left(\partial^a\bar{\phi}\right)
  \left(\partial_b\bar{\phi}\right)
  +\frac{(d-2)}{4(d-1)}
  \bar{\mathcal{R}}^a_{\,\,\,b}\bar{\phi}^2+
  \frac{(d-2)}{4(d-1)}
  \left[
  \left(1-\frac{1}{4\xi}\right)
  \delta^a_b\bar{\nabla}^2-\bar{\nabla}^a\bar{\nabla}_b
  \right]\bar{\phi}^2
  \ ,
\end{equation}
where we have rearranged the standard expression using the equation of motion in Eq. (\ref{eq:12}). From this we can determine the thermal stress tensor from the coincidence limit of the thermal two point functions in Eqs. (\ref{eq:24}) and (\ref{eq:25}) and its derivatives.

The immediate problem with this coincidence limit is that it gives rise to short distance divergences, which we must be regularized. We can do so in the simplest way, by defining the regularized two point function $\bar{G}^{\rm reg}_\beta(X,X')$ according to
\begin{equation}\label{eq:22}
\bar{G} _\beta^{\rm reg}(X,X')=
  \bar{G}_\beta(X,X')-\bar{G}_{\beta=2\pi\alpha}(X,X')\ ,
\end{equation}
where we have chosen the reference temperature as $\beta=2\pi\alpha$ given that it corresponds to the vacuum when mapping back to Minkowski. Using the regularized two point function, we obtain a finite result for the expectation value of the operators in Eq. (\ref{eq:23})
\begin{equation}\label{eq:40}
\langle \bar{\phi}^2 \rangle_\beta=
  \lim_{X'\rightarrow X}
  \bar{G} _\beta^{\rm reg}(X,X')\ ,
  \qquad \qquad
  \langle \left(\partial^a \bar{\phi}\right)
  \left(\partial_b \bar{\phi}\right) \rangle_\beta=
  \lim_{X'\rightarrow X}
  \,\,\partial^{a}\, \partial_{b'}\,
  \bar{G} _\beta^{\rm reg}(X,X')\ ,
\end{equation}
where $\partial_{b'}$ is the derivative with respect to $X'$. Let us first make some observations that will simplify the calculation in Eq. (\ref{eq:44}).

First, notice that the space-time dependence of the thermal two point functions in Eqs. (\ref{eq:24}) and (\ref{eq:25}), is entirely written in terms of $\Delta \eta$ and $\Delta \gamma$. Since taking the coincidence limit is given by $(\Delta \eta,\Delta \gamma)\rightarrow 0$, this means that $\langle \bar{\phi}^2 \rangle_\beta$ will be independent of the space-time coordinates. Thus, the third contribution to the stress tensor in Eq. (\ref{eq:23}), which involves derivatives acting on $\bar{\phi}^2$ drops out.

The first term in Eq. (\ref{eq:23}) is the tricky one, given that when taking the spatial derivatives we must differentiate $\Delta \gamma$, whose complicated spatial dependence is given through Eq. (\ref{eq:16}). The computation can be simplified by noting that since the hyperbolic plane $\mathbb{H}^{d-1}$ is maximally symmetric and the thermal state is by definition homogeneous, there are no preferred spatial directions in the index structure of $\langle \bar{T}^a_{\,\,\,b} \rangle_\beta$. Since the second term in Eq. (\ref{eq:23}) is given by the Ricci tensor that is already homogeneous in space (see Eq. (\ref{eq:51})), all the spatial components of  $\langle \left(\partial^a \bar{\phi}\right)\left(\partial_b \bar{\phi}\right) \rangle_\beta$ are equal. We can then compute the $(u,u)$ component, and use it to infer the rest of them. In other words, the thermal stress tensor can be written as
$$\langle\bar{T}^a_{\,\,\,b}\rangle_\beta=
  {\rm diag}\left(
  -\bar{\mathcal{E}},\bar{p},\dots,\bar{p}
  \right)
  \ ,$$
with the energy density $\bar{\mathcal{E}}$ and  the pressure $\bar{p}$ given by
\begin{equation}\label{eq:28}
\bar{\mathcal{E}}(\beta)=
  \langle \left(\partial_\eta\bar{\phi}\right)
  \left(\partial_\eta\bar{\phi}\right) \rangle_\beta\ ,
  \qquad \quad
  \bar{p}(\beta)=
  \frac{1}{\alpha^2}
  \left[
  \langle\left(\partial_u\bar{\phi}\right)
  \left(\partial_u\bar{\phi}\right)\rangle_\beta
  -\frac{\langle \bar{\phi}^2 \rangle_\beta}
  {d-1}
  \left(\frac{d-2}{2}\right)^2
  \right]\ .
\end{equation}
Notice that we have not used the zero trace condition of the stress tensor to relate the energy and pressure according to $\bar{\mathcal{E}}(\beta)=(d-1)\bar{p}(\beta)$. Such relation is not at all obvious from Eq. (\ref{eq:28}) but will be true when computing explicitly both quantities through Eq. (\ref{eq:40}), using the thermal two point functions in Eqs. (\ref{eq:24}) and (\ref{eq:25}). 

Due to the regularization procedure (\ref{eq:22}), the energy density computed in this way is determined up to an overall constant, which can be fixed from the knowledge of the energy density at any given fixed temperature. In Ref. \cite{Herzog:2015ioa} (see also \cite{Brown:1977sj,Herzog:2013ed}) the thermal energy density at temperature $\beta=2\pi\alpha$ was computed for an arbitrary CFT and shown to be given by
\begin{equation}\label{eq:153}
\bar{\mathcal{E}}(2\pi \alpha)=
  \begin{cases} 
  \,\,\,\,
  \dfrac{4A_d}{d\,\alpha^d{\rm Vol}(S^d)} \ ,
  &\,\, {\rm for\,\,}d{\,\,\rm even} \\
  \quad\qquad \, 0\, \qquad \,\ ,
 & \,\,\,\,\, {\rm for\,\,}d{\,\,\rm odd}\ , \\
 \end{cases}
\end{equation}
where $A_d$ is the coefficient of the Euler density in the trace anomaly (see Ref. \cite{Myers:2010tj} for conventions). Using this, we can fix the undetermined constant and unambiguously determine the thermal energy density.

Before doing so we note that we can also obtain the thermal free energy, defined from the partition function $Z={\rm Tr}(e^{-\beta\bar{H}_\eta})$ as
\begin{equation}\label{eq:68}
\bar{F}(\beta)=
  -\ln\left(Z\right)/\beta=
  \bar{E}(\beta)-\bar{S}(\beta)/\beta\ .
\end{equation}
where $\bar{E}(\beta)$ is the total thermal energy and $\bar{S}(\beta)$ the Von-Neumann entropy of the thermal state. Differentiating with respect to $\beta$ we obtain the first law, which gives an integral expression for $\bar{S}(\beta)$ in terms of $\bar{E}(\beta)$. We can then express the free energy entirely in terms of $\bar{\mathcal{E}}(\beta)$ as
\begin{equation}\label{eq:47}
\bar{F}(\beta)=
  \left[
  \bar{\mathcal{E}}(\beta)-\frac{1}{\beta}
  \int d\beta'
  \beta'\, \frac{d\bar{\mathcal{E}}(\beta')}{d\beta'}
  \right]
  \alpha^{d-1}
  {\rm Vol}(\mathbb{H}^{d-1})+\bar{F}_0\ ,
\end{equation}
where $\bar{F}_0$ is an integration constant. We now proceed to find explicit expressions for $\bar{\mathcal{E}}(\beta)$, $\bar{p}(\beta)$ and $\bar{F}(\beta)$ using Eqs. (\ref{eq:28}) and (\ref{eq:47}).

\subsubsection*{Even space-time}

For even space-time dimensions the calculation is straightforward, since the expression of the two point function in Eq. (\ref{eq:24}) is explicit. However, the result cannot be written for arbitrary $d$, but must be computed for each particular value. For the first few even dimensions we obtain the expressions in Table \ref{table:4}, where we have used Eq. (\ref{eq:153}) to fix the undetermined constant. In every case, we find that the pressure is related to the energy density through ${\bar{\mathcal{E}}(\beta)=(d-1)\bar{p}(\beta)}$, as required by the vanishing trace of the stress tensor, despite that this is not evident from Eq. (\ref{eq:28}). For larger values of $d$ the expressions get more complicated but their qualitative behavior remains unchanged. Using Eq. (\ref{eq:47}) we have also computed the thermal free energy in Table \ref{table:4}.
\begin{table}[]\setlength{\tabcolsep}{10pt}
\centering
\begin{tabular}{|Sc|Sc|Sc|Sc|Sc|}
\hline 
$d$ & 4 & 6 & 8  \\  \hline
$\bar{\mathcal{E}}(\beta)$  &
 $\displaystyle
 \frac{2T^4-1}
 {960\pi^2\alpha^4}$  &
  $\displaystyle
  \frac{20T^6+42T^4-37}
  {120960 \pi^3\alpha^6}$  &
  $\displaystyle
  \frac{168T^8+800T^6+1344T^4-1507}
  {9676800\pi^4\alpha^8}$ 
  \\ \hline
$\dfrac{\bar{F}(\beta)-\bar{F}_0}{{\rm Vol}(\mathbb{H}^{d-1})}$  &
  $-\displaystyle\frac{2T^4+3}
  {2880\pi^2\alpha}
  $  &
  $-\displaystyle
  \frac{4T^6+14T^4+37}
  {120960\pi^3\alpha}
  $  &
  $-\displaystyle
  \frac{24T^8+160T^6+448T^4+1507}
  {9676800\pi^4\alpha}
  $   \\ \hline
\end{tabular}
\caption{Thermal energy and free energy of a scalar field conformally coupled to the hyperbolic background $\mathbb{R} \times \mathbb{H}^{d-1}$, written in terms of the dimensionless temperature $T=2\pi \alpha/\beta$ for the first few even space-time dimensions.}\label{table:4}
\end{table}

\subsubsection*{Odd space-time}

For odd values of $d$, the thermal two point function has an integral expression (\ref{eq:25}) which makes the computations more complicated. However, for $d=3$, we can still write the energy density from Eq. (\ref{eq:28}) in terms of the following integral
$$
\bar{\mathcal{E}}(\beta)=
  \frac{1}
  {4\pi^2\alpha^3\sqrt{2}}
  \int_{0}^\infty
  \frac{dv}{\left(\cosh(v)-1\right)^{5/2}}
  \left[
  \sinh(v)-
  \left(\frac{T\sinh(v/2)}{\sinh(Tv/2)}\right)^4
  \frac{\sinh(Tv)}{T}
  \right]\  .
$$
For several rational values of $T$ the integral can be solved exactly, the most interesting being the zero temperature case where we find 
$$
\bar{\mathcal{E}}(\beta\rightarrow \infty)=
  -\frac{3\zeta(3)}{32\pi^4\alpha^3}\ ,
$$
with $\zeta(z)$ is the Riemann zeta function. For general $T$, we can solve through numerical integration and obtain the plot in Fig. \ref{fig:3}, the red dots corresponding to values of $T$ which allow for exact integration. As required from Eq. (\ref{eq:153}) the expression vanishes for ${\beta=2\pi \alpha}$. Using this in Eq. (\ref{eq:47}) we can also find an integral expression for the free energy
\begin{equation}\label{eq:71}
\begin{aligned}
\bar{F}(\beta)-\bar{F}_0=
  \frac{{\rm Vol}(\mathbb{H}^{2})}{4\pi^2\alpha\sqrt{2}}&
  \int_0^\infty			
  \frac{dv}{\left(\cosh(v)-1\right)^{5/2}}
  \,\times\\  
  &\left[
    \sinh(v)
  -\left(
  \frac{T\sinh(v/2)}{\sinh(Tv/2)}
  \right)^2
  \frac{\sinh^2(v/2)}{(v/2)}
  \left(1+\frac{\sinh(Tv)}{Tv}\right)
  \right]\ ,
\end{aligned}
\end{equation}
that can be solved through numerical integration.

\section{Solution to hyperbolic wave equation}
\label{app:hypeq}

In this Appendix we show how to solve the wave equation (\ref{eq:12}) of a conformally coupled scalar field to the hyperbolic space-time metric $\mathbb{R}\times \mathbb{H}^{d-1}$. Though the solution and its properties can be found in the excellent Ref. \cite{Bander:1965im}, we find it instructive to show its derivation since it is missing from \cite{Bander:1965im}.

It will be convenient to define the coordinate $z=\cosh(u)$, so that the hyperbolic Laplacian in Eq. (\ref{eq:12}) can be easily computed and we find
$$\left[
  (z^2-1)
  \partial_z^2
  +
  (d-1)z
  \partial_z+
  \frac{\Delta_{S^{d-2}}}{(z^2-1)}
  -
  \alpha^2\partial_\eta^2+
  \left(\frac{d-2}{2}\right)^2
  \right]
  \phi(x)=0\ ,$$
where $\Delta_{S^{d-2}}$ is the Laplacian of the unit sphere $S^{d-2}$. Considering the ansatz in Eq. (\ref{eq:31}) and using that the eigenvalues of the spherical harmonics are given by $-\ell(\ell+d-3)$, we find the following differential equation for $f_{w,\ell}(z)$
\begin{equation}\label{eq:32}
(z^2-1)
  f''(z)
  +
  (d-1)zf'(z)+
  \left[
  w^2+
  \left(\frac{d-2}{2}\right)^2
  -
  \frac{\ell(\ell+d-3)}{(z^2-1)}
  \right]
  f(z)=0\  .
\end{equation}
To solve this, we first consider the behavior of $f(z)$ when $z\sim \pm 1$, where it is easy to see that in this limit the solution is given by $f(z)\sim (z\mp 1)^{\ell/2}$. It is then useful to write the function as $f(z)=(z^2-1)^{\ell/2}h(z)$, and plug this into Eq. (\ref{eq:32}), so that we obtain the following differential equation for $h(z)$
\begin{equation}\label{eq:33}
(1-z^2)h''(z)-
  2\left(\frac{d-1}{2}+\ell\right)z\,h'(z)-\left[
  w^2+\left(\frac{d-2}{2}+\ell\right)^2
  \right]h(z)=0\ .
\end{equation}
The solution of this equation splits depending on whether $d$ is even or odd.

\subsubsection*{Even space-time}

For even values of $d$ it is convenient to define the factor $n=(d-2)/2+\ell$, that is an integer larger or equal than one. For $n=0$, Eq. (\ref{eq:33}) becomes
\begin{equation}\label{eq:34}
(1-z^2)h_{n=0}''(z)-
  z\,h_{n=0}'(z)-w^2h_{n=0}(z)=0\ .
\end{equation}
Writing this equation in terms of the $u$ coordinate  $z=\cosh(u)$, we obtain a harmonic oscillator equation with frequency $w$. We can then write the solution for the $n=0$ case as $h_{n=0}(z)\propto \cos(wu)$. For arbitrary values of $n$, we notice that if we take the $n$-th derivative of Eq. (\ref{eq:34}) with respect to $z$ we get
$$(1-z^2)\frac{d^2}{dz^2} \left[h^{(n)}_{n=0}(z)\right]
  -(1+2n)z\,\frac{d}{dz} \left[h_{n=0}^{(n)}(z)\right]
  -\left(w^2+n^2\right)
  h_{n=0}^{(n)}(z)=0\ ,$$
where $h_{n=0}^{(n)}(z)$ is the $n$-th derivative of $h_{n=0}(z)$. We recognize this equation as the one we started from in Eq. (\ref{eq:33}). This means that the general solution in the $u$ variables is given by taking derivatives of the simple $n=0$ solution
\begin{equation}\label{eq:37}
f_{w,\ell}(u)=
  N\sinh^\ell(u)
  \left(\frac{d}{d\cosh(u)}\right)^{\frac{d-2}{2}+\ell}
  \cos(wu)\ ,
\end{equation}
with $N$ a normalization constant.

\subsection*{Odd space-time}

For the case of odd space-time, it is convenient to write Eq. (\ref{eq:33}) in terms of the integer $m=(d-3)/2+\ell$ and the parameter $\nu=iw-1/2$, so that the differential equation becomes
\begin{equation}\label{eq:36}
(1-z^2)h''(z)-
  2(m+1)z\,h'(z)+(\nu-m)(\nu+m+1)h(z)=0\ .
\end{equation}
For the $m=0$ case, we recognize the differential equation as the ordinary Legendre equation with parameter $\nu$ \cite{Gradshteyn,Trans}
\begin{equation}\label{eq:35}
(1-z^2)h_{m=0}''(z)-
  2z\,h_{m=0}'(z)+\nu(\nu+1)h_{m=0}(z)=0\ .
\end{equation}
Since we want the solution to be regular when $z\sim 1$, we consider $h_{m=0}(z)\propto P_{iw-1/2}(z)$. For arbitrary $m \in \mathbb{Z}_{\ge 0}$, we take the $m$-th derivative of Eq. (\ref{eq:35}) with respect to $z$, and find
$$(1-z^2)
  \frac{d^2}{dz^2}\left[
  h^{(m)}_{m=0}(z)\right]-
  2(m+1)z\,
  \frac{d}{dz}\left[h_{m=0}^{(m)}(z)\right]+(\nu-m)(\nu+m+1)h^{(m)}_{m=0}(z)=0\ ,$$
where $h^{(m)}_{m=0}(z)$ is the $m$-th derivative. This is again the same equation we started from (\ref{eq:36}), which means that the general solution can be written in the $u$ coordinate by taking derivatives of the $m=0$ solution
\begin{equation}\label{eq:38}
f(u)=N
  \sinh^\ell(u)
  \left(
  \frac{d}{d\cosh(u)}
  \right)^{\frac{d-3}{2}+\ell}
  P_{iw-1/2}(\cosh(u))\ .
\end{equation}

\subsubsection*{Completeness and orthogonality}

In Ref. \cite{Bander:1965im} it was shown that these solutions form an orthonormal set
$$
  \int_0^{\infty}
  du\,\left(\sinh(u)\right)^{d-2}
  f_{w,\ell}(u)
  f^*_{w',\ell}(u)
  =\delta(w-w')
  \ ,$$
as long as the normalization constant $N$ in Eqs. (\ref{eq:37}) and (\ref{eq:38}) is taken as
$$N^2_{\rm even}=
  \frac{2}{\pi}
  \prod_{q=0}^{\frac{d-4}{2}+\ell}
  \frac{1}{w^2+q^2}=
  \frac{2\left|\Gamma\big(
  (d-2)/2+\ell+iw
  \big)\right|^{-2}}{w\sinh(\pi w)}
  \ ,$$
$$N^2_{\rm odd}=
  w\tanh(\pi w)
  \prod_{q=0}^{\frac{d-5}{2}+\ell}
  \frac{1}{w^2+(q-1/2)^2}=
  \left(\frac{\pi w}{w^2+1/4}\right)
  \frac{\sinh(\pi w)}{\cosh^2(\pi w)}
  \left|\Gamma\big((d-4)/2+\ell+iw\big)\right|^{-2}\ .
  $$
Moreover, the following completeness relation was shown to be true for the mode solutions $H_{\vec{k}}(X)$ in Eq. (\ref{eq:31})
$$\sum_{\ell,m}
  \int_0^{\infty}dw\,
  H_{\vec{k}}(X)
  H^*_{\vec{k}}(X')
  \Big|_{\tau=\tau'}=
  \delta_{\rm hyp.}(X,X')\ ,$$
where $\delta_{\rm hyp.}(X,X')$ is the Dirac delta in the unit hyperboloid. These relations are crucial for the derivation of the commutation relations of $\bar{\phi}(X)$ and $\bar{\pi}(X)$.

\section{Quasi-local stress tensor of hyperbolic black hole}
\label{app:quasi}

In this Appendix we obtain the thermal energy density in the hyperbolic space-time $\mathbb{R}\times \mathbb{H}^{d-1}$ for a strongly coupled CFT using the AdS/CFT correspondence. We do so by computing the quasi-local stress tensor \cite{Brown:1992br,Balasubramanian:1999re} of the appropriate black hole solutions for Einstein and Gauss-Bonnet gravity theories.

\subsection{Einstein gravity}

Let us start by considering Einstein gravity, which has a bulk action given by 
\begin{equation}\label{eq:26}
I_{\rm bulk}=
  \frac{1}{2\ell_p^{d-1}}
  \int d^{d+1}x\sqrt{-g}
  \left[
  \frac{d(d-1)}{L^2}+R
  \right]\ ,
\end{equation}
where $\ell_p$ is Planck's length, $L$ the AdS radius and $R$ the Ricci scalar. The generalized central charge of the dual field theory $a_d^*$ defined in Eq. (\ref{eq:120}), is computed from the pure AdS solution according to \cite{Myers:2010tj}
\begin{equation}\label{eq:79}
a_d^*=-\frac{{\rm Vol}(S^{d-1})L^{d+1}}{2d}
  \mathcal{L}\big|_{\rm AdS}=
  \frac{1}{2}{\rm Vol}(S^{d-1})
  \left(\frac{L}{\ell_p}\right)^{d-1}\ ,
\end{equation}
where $\mathcal{L}\big|_{\rm AdS}$ is the Lagrangian density in Eq. (\ref{eq:26}) evaluated at the pure AdS solution. The thermal state in the background $\mathbb{R}\times \mathbb{H}^{d-1}$ is dual to a black hole with a hyperbolic horizon
\begin{equation}\label{eq:155}
ds^2=
  -V(\rho)\left(L/\alpha\right)^2
  d\eta^2+
  \frac{d\rho^2}{V(\rho)}+
  \rho^2dH_{d-1}^2\ ,
\end{equation}
where $dH_{d-1}$ is the line element of a unit hyperbolic plane and the radial function $V(\rho)$ is fixed from the equations of motion according to
\begin{equation}\label{eq:127}
V(\rho)=-1+\left(\rho/L\right)^2-\mu/\rho^{d-2}\ ,
\end{equation}
with $\mu$ an integration constant. We have rescaled the time coordinate $\eta$ in Eq. (\ref{eq:155}) so that we recover the hyperbolic $d$-dimensional metric with radius $\alpha$ at the AdS boundary $\rho\rightarrow +\infty$. This space-time has a causal horizon at $\rho=\rho_+$ defined from $g^{\rho \rho}(\rho_+)=0$. From this constraint we can write the integration constant $\mu$ in terms of $\rho_+$ as
\begin{equation}\label{eq:72}
\frac{\mu }{L^{d-2}}=
  x_+^{d-2}\left(
  x_+^2-1
  \right)\ ,
  \qquad \qquad
  x_+=\rho_+/L\ .
\end{equation}
The associated inverse temperature $\beta$ can be easily computed from the surface gravity $\kappa$ 
\begin{equation}\label{eq:121}
\beta=\frac{2\pi }{\kappa}=
  2\pi \alpha
  \left(\frac{2x_+}{x_+^2d-(d-2)}\right)
  \qquad \Longrightarrow \qquad
  x_+=
  \frac{T+\sqrt{T^2+d(d-2)}}{d}\ ,
\end{equation}
where we have defined the dimensionless temperature $T=2\pi\alpha/\beta$.

We now wish to compute the quasi-local stress tensor associated to this solution. To do so, we must first introduce a radial cut-off $\rho_0$ and consider the following regularized on-shell action
$$I_{\rm on-shell}=
  \frac{1}{2\ell_p^{d-1}}
  \int d^{d+1}x\sqrt{-g}
  \left[
  \frac{d(d-1)}{L^2}+R
  \right]+
  \frac{1}{\ell_p^{d-1}}\int_{\rho=\rho_0} d^dX\sqrt{-h}K-
  I_{\rm ct}\left[h_{ab}\right]\ ,$$
where the coordinates $X^a$ parametrize the boundary $\rho=\rho_0$ and $K$ is the trace of the extrinsic curvature associated to the boundary with metric $h_{ab}$. The second term is the Gibbons-Hawking-York (GHY) boundary term that has to be considered for the variational principle to be well defined. The third term is the counter term needed to regulate divergences and yield a finite result \cite{Hyun:1998vg,Balasubramanian:1999re,Emparan:1999pm}. The structure of this term depends on the space-time dimensions. From the on-shell action we can define the quasi-local stress tensor as \cite{Brown:1992br}
\begin{equation}\label{eq:81}
T_{ab}=-\frac{2}{\sqrt{-h}}
  \frac{\delta I_{\rm on-shell}}{\delta h^{ab}}\ .
\end{equation}
For $d\le 4$ this can be computed and gives \cite{Balasubramanian:1999re}
\begin{equation}\label{eq:156}
\ell_p^{d-1}T^a_{\,\,\,b}=
  \left(K\delta^a_b-K^a_{\,\,\,b}\right)-
  \left(\frac{d-1}{L}\delta^a_{\,\,\, b}-
  \Theta(d-5/2)\frac{L}{d-2}G^a_{\,\,\,b}\right)\ .
\end{equation}
where $G_{ab}=\mathcal{R}_{ab}-h_{ab}\mathcal{R}/2$ is the Einstein tensor of the boundary metric. The first term comes from the GHY contribution while the second from the counter term. Computing this explicitly we can easily guess the general expression for arbitrary values of $d$ by requiring that the correct ADM mass is recovered when computing the conserved charge under time translations. The end result is given by
$$T^a_{\,\,\,b}=
  \frac{a_d^*}{{\rm Vol}(S^{d-1})}
  \left[
  x_+^{d-2}\left(
  x_+^2-1
  \right)+c_d
  \right]
  {\rm diag}
  \left(
  1-d,1,\dots,1
  \right)\rho_0^{-d}
  +\mathcal{O}(\rho_0^{-d-1})\ ,$$
where we have used Eqs. (\ref{eq:79}) and (\ref{eq:72}). The constant term $c_d$ is a Casimir type contribution that is non vanishing only for $d$ even and has been computed in Ref. \cite{Emparan:1999pm}
$$
c_d=
  \begin{cases}
  \,\,\, 
  \displaystyle
  \frac{4{\rm Vol}(S^{d-1})}{d(d-1){\rm Vol}(S^d)}
  \,\ ,
  & {\rm for\,\,d\,\,even} \vspace{6pt}\\
  \quad
  \qquad \quad
  0
  \qquad \quad \,\,\ ,
  &  {\rm for\,\,d\,\,odd}\ , \\
 \end{cases}
$$
where ${\rm Vol}(S^{d-1})=2\pi^{d/2}/\Gamma(d/2)$. From this we can compute the expectation value of the stress tensor operator $\bar{T}^a_{\,\,\,b}$ in the thermal state $\bar{\rho}(\beta)$ in the hyperbolic background as
\begin{equation}\label{eq:157}
{\rm Tr}\left(
  \bar{\rho}(\beta)\,
  \bar{T}^a_{\,\,\,b}
  \right)=
  \lim_{\rho_0\rightarrow +\infty}
  \left(\frac{\rho_0}{\alpha}\right)^d
  T^a_{\,\,\,b}\ .
\end{equation}
Doing so, we can read off the energy density as
\begin{equation}\label{eq:124}
\bar{\mathcal{E}}(\beta)=
  \frac{a_d^*(d-1)}{{\rm Vol}(S^{d-1})\alpha^d}
  g(\beta)^{d-2}\left(
  g(\beta)^2-1
  \right)+\bar{\mathcal{E}}(2\pi \alpha)
  \ ,
\end{equation}
where we have defined $g(\beta)=x_+$ in Eq. (\ref{eq:121}) for notation convenience. The first term vanishes when $\beta=2\pi \alpha$, so that the energy density in the hyperbolic space-time at $\beta=2\pi \alpha$ is given by the second term
\begin{equation}\label{eq:123}
\bar{\mathcal{E}}(2\pi \alpha)=
  \begin{cases}
  \,\,\, 
  \displaystyle
  \frac{4a_d^*}{d\,\alpha^d{\rm Vol}(S^d)}
  \,\ ,
  & {\rm for\,\,d\,\,even} \vspace{6pt}\\
  \quad
  \qquad 
  0
  \qquad  \,\,\,\ ,
  &  {\rm for\,\,d\,\,odd}\ , \\
 \end{cases}
\end{equation}
in agreement with the CFT result in Eq. (\ref{eq:153}). 

For odd space-time dimensions $\bar{\mathcal{E}}(\beta)$ is negative for $\beta>2\pi \alpha$. For even $d$ we have the extra Casimir contribution which is always positive. For $d>4$ this gives a new smaller range of $\beta$ for which $\bar{\mathcal{E}}(\beta)$ is negative, while for $d=2,4$ it is always positive. This behavior is true for both the mass of the black hole and the energy of the thermal state in $\mathbb{R}\times \mathbb{H}^{d-1}$.

\subsection{Gauss-Bonnet gravity}

We can now consider higher curvature corrections to the bulk action (\ref{eq:26}) by adding a Gauss-Bonnet term, which will modify the equations of motion for $d\ge 4$. Since the calculation becomes more complex, we consider the $d=4$ case that will capture most of the relevant features. In this case, the bulk action is given by
\begin{equation}\label{eq:80}
I_{\rm bulk}=
  \frac{1}{2\ell^{3}_p}
  \int d^{5}x\sqrt{-g}
  \left[
  \frac{12}{ L^2}+
  R+
  \frac{\lambda L ^2}{2}\mathcal{X} _4
  \right]\ ,
\end{equation}
where
$$\mathcal{X}_4=
  R_{\mu \nu \rho \sigma}R^{\mu \nu \rho \sigma}-
  4R_{\mu \nu}R^{\mu \nu}+R^2\ .$$ 
Pure AdS will be a solution to the equations of motion as longs as its radius $\tilde{L}$ is given by
$$\tilde{L}=L/f_\infty\ ,
  \qquad \qquad
  f_\infty=\frac{1-\sqrt{1-4\lambda}}{2\lambda}\ .$$
From this we can compute the value of the generalized central charges of the boundary CFT, given by the coefficients of the anomalous terms appearing in the trace of the stress tensor. In this case, the two coefficients $a$ and $c$ will not be the same (as they are for Einstein gravity) and can be found to be \cite{Myers:2010tj}
\begin{equation}\label{eq:9}
a_4^*=
  \pi^2
  \left(
  1-6\lambda f_\infty
  \right)
  \left(\frac{\tilde{L}}{\ell_p}\right)^3
  \ ,
  \qquad \qquad
  c=\pi^2
  (1-2\lambda f_\infty)
  \left(\frac{\tilde{L}}{\ell_p}\right)^3\ .
\end{equation} 
The thermal state in the boundary theory will be described by a black hole solution with hyperbolic horizon, that is given by \cite{Cai:2001dz}
\begin{equation}\label{eq:82}
ds^2=
  -\left(
  \frac{\rho^2}{L^2}f(\rho)-1
  \right)(\tilde{L}/\alpha)^2d\eta^2+
  \frac{d\rho^2}
  {\left(
  \frac{\rho^2}{L^2}f(\rho)-1
  \right)}
  +\rho^2dH_{3}^2\ ,
\end{equation}
where
$$f(\rho)=
  \frac{1}{2\lambda }
  \left(
  1-\sqrt{1+\frac{4\lambda L^2  }{\rho^4}\mu-4\lambda }\,
  \right)\ ,
  \qquad \qquad
  \lim_{\rho \rightarrow +\infty}f(\rho)=f_\infty\ ,
  $$
and $\mu$ is an integration constant. Taking the boundary limit $\rho\rightarrow +\infty$ we recover the boundary metric $\mathbb{R}\times \mathbb{H}^{3}$ of the CFT. From the horizon radius $\rho=\rho_+$ defined from $g^{\rho \rho}(\rho_+)=0$, we can write the integration constant $\mu$ as
\begin{equation}\label{eq:158}
\frac{\mu}{L^2}=
  x_+^4-
  x_+^2+
  \lambda\ ,
  \qquad \qquad
  x_+=\rho_+/L\ .
\end{equation}
The temperature is computed from the surface gravity $\kappa$ evaluated at the horizon and it gives
\begin{equation}\label{eq:10}
\beta=\frac{2\pi}{\kappa}=
  2\pi \alpha f_\infty
  \left(
  \frac{x_+^2-2\lambda}
  {2x_+^3-x_+}
  \right)
  \ .
\end{equation}
Though we could invert this expression to find $x_+$ as a function of $\beta$, we prefer not to do it explicitly since it involves solving for the root of a cubic polynomial.

The quasi-local stress tensor is computed again from Eq. (\ref{eq:81}) but considering the generalizations of the GHY and counter terms appropriate to Gauss-Bonnet gravity \cite{Myers:1987yn,Brihaye:2008kh}. Doing so, the stress tensor is found to be \cite{Brihaye:2008kh}
$$\ell_p^3 T^a_{\,\,\,b}=
  \left(K\delta^a_{\,\,\,b}-K^a_{\,\,\,b}\right)
  +
  \lambda L^2
  \left(
  \frac{1}{3}
  Q\delta^a_{\,\,\,b}-
  Q^a_{\,\,\,b}
  \right)-
  \frac{1}
  {\tilde{L}}
  \left(\frac{f_\infty+2}{f_\infty}\right)
  \delta^a_{\,\,\,b}+
  \frac{\tilde{L}}{2}
  \left(\frac{3f_\infty-2}{f_\infty}\right)
  G^a_{\,\,\,b}
  \ ,$$
where
$$Q^{a}_{\,\,\,b}=
  2KK^a_{\,\,\,c}K^c_{\,\,\,b}-
  2K^a_{\,\,\,c}K^{cd}K_{db}+
  K^a_{\,\,\,b}
  \big(K_{cd}K^{cd}-K^2\big)+
  2K\mathcal{R}^a_{\,\,\,b}+
  \mathcal{R}K^a_{\,\,\,b}+
  2K^{cd}\mathcal{R}^a_{\,\,\,cdb}-
  4\mathcal{R}^a_{\,\,\,c}K^c_{\,\,\,b}\ .$$
The term involving the $Q^a_{\,\,\,b}$ tensor comes from the Gauss-Bonnet contribution to the GHY term. In the limit of $\lambda \rightarrow 0$ we recover the Einstein expression (\ref{eq:156}) evaluated at $d=4$. Using the hyperbolic black hole metric (\ref{eq:82}), we can compute this tensor and find
$$T^a_{\,\,\,b}=
  \frac{\tilde{L}^3}{2\ell_p^3}
  \left[
  \frac{\mu }{\tilde{L}^2} +
  \frac{\left(1-6\lambda f_\infty\right)}{4}
  \right]
  {\rm diag}\left(
  -3,1,1,1
  \right) \rho_0^{-4} 
  +\mathcal{O}(\rho_0^{-5})\ .$$
The parameters $\lambda$ and $f_\infty$ can be written in terms of the CFT central charges $a$ and $c$ using Eq. (\ref{eq:9})
\begin{equation}\label{eq:159}
\lambda=
  \frac{
  (5c-a)(c-a)}
  {4(3c-a)^2}
  \qquad \Longrightarrow \qquad
  f_\infty=\frac{2(3c-a)}{5c-a}\ .
\end{equation}
Taking the boundary limit as in Eq. (\ref{eq:157}) we find that the thermal energy density of the CFT is given by
$$\bar{\mathcal{E}}(\beta)=
  \frac{3a}{{\rm Vol}(S^3)\alpha^4}
  \left(\frac{x_+^4-x_+^2+\lambda}
  {f_\infty^{-1}-6\lambda}\right)
  +
  \bar{\mathcal{E}}(2\pi \alpha)\ .
  $$
where we have used Eq. (\ref{eq:158}). Apart from the $\beta=2\pi\alpha$ case, where the first term vanishes and the energy density is given in Eq. (\ref{eq:123}), the result is different from the one obtained from the Einstein calculation (\ref{eq:124}). The parameter $x_+$ can be written explicitly in terms of $\beta$ and the central charges by inverting Eq. (\ref{eq:10}) and using Eqs. (\ref{eq:159}). In the zero temperature limit the energy density simplifies and is given by
\begin{equation}\label{eq:125}
\bar{\mathcal{E}}(\beta\rightarrow \infty)=
  -\frac{3c^2}{2\pi^2\alpha^4(5c-a)}+
  \bar{\mathcal{E}}(2\pi \alpha)\ .
\end{equation}
The values of $c$ and $a$ are not arbitrary but are constrained according to Eq. (\ref{eq:160}). When $a=c$ the vacuum energy density in the hyperbolic space vanishes, in agreement with the Einstein gravity result in Eq. (\ref{eq:124}).

\section{Surface energy in separable states}
\label{app:Juan}

\footnote{We thank Juan Hernandez for helpful discussions which resulted in this Appendix and an anonymous referee whose inquires resulted in its inclusion in this paper.}Consider a QFT in $d$-dimensional Minkowski space-time in a state described by any density operator $\rho$. The Cauchy surface $t=0$ can be split in half by the infinite plane $x=0$, where we write the spatial coordinates as $\vec{x}=(x,\vec{y})$. Assuming that the total Hilbert space admits a tensor product structure in terms of the left and right degrees of freedom $\mathcal{H}=\mathcal{H}_L\otimes \mathcal{H}_R$, we can define the reduced density operators $\rho_{L/R}$ by tracing over the complementary regions. These operators satisfy the following property
\begin{equation}\label{eq:123b}
{\rm Tr}_{\mathcal{H}_{L/R}}\left(
  \rho_{L/R}\, \mathcal{O}_{L/R} \right)=
  {\rm Tr}_{\mathcal{H}}\left(
  \rho \,\mathcal{O}_{L/R}
  \right)=\langle \mathcal{O}_{L/R} \rangle_\rho
  \ ,
\end{equation}
where $\mathcal{O}_{L/R}$ is any operator localized in the left and right side respectively. From this we can build a separable state that is globally defined in $\mathcal{H}$ as $\rho_{\rm sep}=\rho_L\otimes \rho_{R}$, whose expectation value on any local observable $\mathcal{O}_{L/R}$ is given by
\begin{equation}\label{eq:125b}
{\rm Tr}_{\mathcal{H}}\left(
  \rho_{\rm sep}\,\mathcal{O}_{L/R}
  \right)=
  \langle \mathcal{O}_{L/R} \rangle_\rho\ .
\end{equation} 
The question we want to address is what happens for an observable localized precisely at the entangling plane, \textit{i.e.} $\lim_{x\rightarrow 0}\langle\mathcal{O}(x,\vec{y})\rangle_{\rho_{\rm sep}}$. In particular, we are interested in its energy. For definiteness, let us consider a scalar field whose action is given by
$$S_{\rm scalar}=-\int d^dx \left[
  \frac{1}{2}\left(\partial_\mu \phi\right)
  \left(\partial^\mu\phi\right)+V(\phi)
  \right]\ ,$$
where $V(\phi)$ is an arbitrary function of $\phi$ with a minimum at $\phi=0$. From the conjugate momentum $\pi(t,\vec{x})=\partial_t\phi(t,\vec{x})$ we can easily compute the hamiltonian density which gives the following expression for $T_{tt}(t,\vec{x})$
\begin{equation}\label{eq:124b}
T_{tt}(t,\vec{x})= 
  \frac{1}{2}\pi^2(t,\vec{x})+
  \frac{1}{2}\big(\vec{\nabla}\phi(t,\vec{x})\big)^2
  +V\big(\phi(t,\vec{x})\big)\ ,
\end{equation}
where $\vec{\nabla}$ gives the spatial derivatives in $(x,\vec{y})$. Since it contains quadratic and higher product of of the fields evaluated at coincident points, this operator must be regularized by a normal ordering procedure
$$\normord{T_{tt}(t,\vec{x})}=
  T_{tt}(t,\vec{x})-
  \bra{0_M}T_{tt}(t,\vec{x})\ket{0_M}\ ,$$
where $\ket{0_M}$ is the Minkowski vacuum state. This subtraction ensures the cancellation of the UV divergences for ordinary states built from the vacuum.

To compute the energy density of the separable state at the entangling surface $(t,x,\vec{y})=(0,0,\vec{y})$ we discretize the $x$ coordinate $x_n=\epsilon(2n-1)$ with $n \in \mathbb{Z}$, and then take the limit $\epsilon \rightarrow 0$. For an arbitrary power of the field $\phi$ we define its expectation value on $\rho_{\rm sep}$ at the entangling surface as
$${\rm Tr}_\mathcal{H}\left(
  \rho_{\rm sep}\,\phi^k(0)
  \right)\equiv
  \lim_{\epsilon \rightarrow 0}
  {\rm Tr}_\mathcal{H}\left(
  \rho_{\rm sep}\,\phi^k(-\epsilon)
  \right)=
  \lim_{\epsilon \rightarrow 0}
  {\rm Tr}_{\mathcal{H}_L}\left(
  \rho_L\,\phi^k(-\epsilon)
  \right)
  {\rm Tr}_{\mathcal{H}_R}\left(
  \rho_{R}
  \right)\ ,$$
where we are omitting the coordinates $t$ and $\vec{y}$ in the dependence of $\phi$ for notation convenience. Using Eq. (\ref{eq:123b}) we can write both factors in terms of the expectation value of $\rho$. Since this global state has no problem at the surface $x=0$, the $\epsilon\rightarrow 0$ limit can be taken directly and we find
\begin{equation}\label{eq:126b}
{\rm Tr}_\mathcal{H}\left(
  \rho_{\rm sep}\,\phi^k(0)
  \right)\equiv
  \lim_{\epsilon \rightarrow 0}
  {\rm Tr}_\mathcal{H}\left(
  \rho_{\rm sep}\,\phi^k(-\epsilon)
  \right)=
  \langle\phi^k(0)
  \rangle_\rho
  \ .
\end{equation}
This means that for arbitrary powers of $\phi$ there is no additional surface contribution and Eq. (\ref{eq:125b}) still holds at $(t,x,\vec{y})=(0,0,\vec{y})$. Though this is still true for derivatives in the $\vec{y}$ direction, we must be careful when considering derivatives across the entangling surface, in the $x$ direction. In this case we must also discretize the derivative so that the expectation value of $(\partial_x\phi)^2$ becomes
$${\rm Tr}_\mathcal{H}\left(
  \rho_{\rm sep}
  \left(\partial_x\phi(0)\right)^2
  \right)\equiv
  \lim_{\epsilon \rightarrow 0}
  {\rm Tr}_\mathcal{H}\left[
  \rho_{\rm sep}
  \left(
  \frac{\phi(\epsilon)-\phi(-\epsilon)}{2\epsilon}\right)^2
  \right]\ .
\footnote{One might also be worried about the time derivative of the field that is hidden in the conjugate momentum $\pi(t,\vec{x})=\partial_t\phi(t,\vec{x})$. It can be treated in the same way by going to null coordinates $u_\pm=x\pm t$ and applying this same procedure to each direction.}
  $$
The key point is that when expanding the square we will encounter product of the fields in opposite sides of the entangling surface, whose expectation value will factorize in the separable state $\langle \phi(\epsilon) \rangle_\rho\langle \phi(-\epsilon)\rangle_\rho$. This means that in order to recover $\langle\left(\partial_x\phi(0)\right)^2\rangle_\rho$ on the right hand side, we must add additional terms in which the expectation value does not factorize. Working this out, the additional term can be written as
\begin{equation}\label{eq:127b}
{\rm Tr}_\mathcal{H}\left(
  \rho_{\rm sep}
  \left(\partial_x\phi(0)\right)^2
  \right)=
  \langle (\partial_x \phi(0))^2 \rangle_\rho
  +
  \lim_{\epsilon \rightarrow 0}
  \frac{C(\rho;\epsilon,\vec{y})}{2\epsilon^2}\ ,
\end{equation}
where
$$2C(\rho;\epsilon,\vec{y})=
  \langle 
  \big(
  \phi(\epsilon)\phi(-\epsilon)\rangle_\rho-
  \langle \phi(\epsilon) \rangle_\rho
  \langle \phi(-\epsilon) \rangle_\rho\big)+
  \big(
  \langle\phi(-\epsilon)\phi(\epsilon)
  \rangle_\rho-
  \langle \phi(-\epsilon) \rangle_\rho
  \langle \phi(\epsilon) \rangle_\rho
  \big)\ .
  $$
The function $C(\rho;\epsilon,\vec{y})$ precisely captures the correlations of $\rho$ across the entangling surface. Moreover, in the limit of $\epsilon \rightarrow 0$ it is non-negative since each term can be written as $\langle \mathcal{O}^2\rangle_\rho-\langle \mathcal{O} \rangle_\rho^2=\langle\left(\mathcal{O}-\langle \mathcal{O} \rangle_\rho\right)^2\rangle_\rho\ge 0$. When taking this limit the second term in Eq. (\ref{eq:127b}) diverges from the $1/\epsilon^2$ factor and from the coincidence limit of the two point function. Notice that applying normal ordering on $(\partial_x\phi)^2$ does not cancel either of these divergences.

Using Eqs. (\ref{eq:126b}) and (\ref{eq:127b}) we can easily compute the energy density of the separable state as
$$\langle \normord{T_{tt}(\vec{x})} \rangle_{\rho_{\rm sep}}=
  \langle \normord{T_{tt}(\vec{x})} \rangle_\rho
  +
  \begin{cases}
  \displaystyle
  \qquad \quad \,\,\,\,  0 \,\,\,\,
  \qquad  \,\,\,\ ,
  \qquad x\neq (0,\vec{y})\\
  \displaystyle
  \,\,\,
  \lim_{\epsilon \rightarrow 0}
  \frac{C(\rho;\epsilon,\vec{y})}{2\epsilon^2}
  \quad\ ,
  \qquad
  \vec{x}=(0,\vec{y})\ .
  \end{cases}$$
We explicitly see there is a surface contribution to the energy density which diverges to positive infinity in the continuum limit. Integrating over all space we can also compute the total energy. Assuming that the initial global state $\rho$ is homogeneous (a thermal state for instance), the term $C(\rho;\epsilon,\vec{y})$ is independent of $\vec{y}$ and we find 
$$E(\rho_{\rm sep})=
  E(\rho)+
  {\rm Vol}(\mathbb{R}^{d-1})
  \lim_{\epsilon \rightarrow 0}
  C(\rho;\epsilon)
  \ .$$
The factor proportional to the spatial volume $\mathbb{R}^{d-1}$ comes from the integral in $\vec{y}$ and a remaining factor $\epsilon^{-1}$. This divergence is not surprising since it is expected for a homogeneous state. On the other hand, the divergence of $C(\rho;\epsilon)$ is induced by UV correlations and shows that in order to construct a separable state in the continuum QFT we require an infinite amount of positive energy localized around the entangling surface.

\bibliographystyle{JHEP}
\bibliography{sample}

\end{document}